\documentclass[10pt, a4paper, british, final]{article}

\pdfoutput=1

\usepackage{\jobname}

\begin{document}
\listoftodos
\pagebreak
\maketitle

\begin{abstract}
\textbf{Purpose:}
The purpose of this work is to apply a recently proposed constitutive model for mechanically induced martensitic transformations to the prediction of transformation loci. Additionally, this study aims to elucidate if a stress-assisted criterion can account for transformations in the so-called strain-induced regime.

\textbf{Design/methodology/approach:}
The model is derived by generalising the stress-based criterion of \citet{Patel1953}, relying on lattice information obtained using the Phenomenological Theory of Martensite Crystallography. Transformation multipliers (cf. plastic multipliers) are introduced, from which the martensite volume fraction evolution ensues. The associated transformation functions provide a variant selection mechanism. Austenite plasticity follows a classical single crystal formulation, to account for transformations in the strain-induced regime. The resulting model is incorporated into a fully-implicit RVE-based computational homogenisation finite element code.

\textbf{Findings:}
Results show good agreement with experimental data for a meta-stable austenitic stainless steel. In particular, the transformation locus is well reproduced, even in a material with considerable slip plasticity at the martensite onset, corroborating the hypothesis that an energy-based criterion can account for transformations in both stress-assisted and strain-induced regimes.

\textbf{Originality/value:}
A recently developed constitutive model for mechanically induced martensitic transformations is further assessed and validated. Its formulation is fundamentally based on a physical metallurgical mechanism and derived in a thermodynamically consistent way, inheriting a consistent mechanical dissipation. This model draws on a reduced number of phenomenological elements and is a step towards the fully predictive modelling of materials that exhibit such phenomena.
\end{abstract}

\keywords{\thekeywords}

\textbf{Article classification:} Research paper

\section{Introduction}

Mechanically induced martensitic phase transformations play a crucial role in the overall mechanical behaviour of many industrially important materials, such as meta-stable austenitic stainless steels and TRIP (\emph{TRansformation Induced Plasticity}) steels \citep{Roters2010}. From a modelling perspective, they pose significant challenges stemming from a complex interplay between a variety of physical mechanisms at the crystal lattice scale. These include, for instance, volumetric expansions across multiple transformation systems and their interactions with crystallographic slip in both the austenitic and martensitic phases.

Due to their inherent microstructural complexity, a phenomenological approach to the constitutive modelling of these materials often falls short of appropriately predicting their mechanical behaviour. The development of robust models relying on a reduced number of phenomenological elements remains largely an open task for this class of materials. In this context, multi-scale models can help elucidating the relationship between microscopic properties and the corresponding macroscopic material behaviour, enabling the simulation of engineering-scale problems that account for the underlying micro-structure in a predictive fashion.

Much research has been dedicated to predicting the crystallographical features of martensitic transformations; the currently most accepted theory is known as the \emph{Phenomenological Theory of Martensite Crystallography} (PTMC), developed in independently by \citet{Wechsler1953} and \citet{Bowles1954}. While it has been extended since its inception\footnote{For a detailed historical account of competing theories formulated both before and after the PTMC, see \cite{Kelly2012}.}, the theory has nevertheless stood the test of time and still is the base of many widely used constitutive models. The PTMC is thus named because it does not aim to explain the mechanisms behind the transformation; instead, it merely predicts the observed crystallographic features of the martensitic transformations in many materials: orientation relationships, habit planes, shape deformation and other geometrical features. The theory was later refined by \citet{Ball1987}, who obtained the same predictions from an energy minimisation perspective.

\citet{Patel1953} observed that mechanical loadings have an influence on the transformation starting temperature. Their proposed explanation for this phenomenon hinged on the work done between applied stresses and transformation displacements. From that, they derived a simple transformation criterion -- limited to uniaxial stress states and small strains -- based on the mechanical contribution to the total thermodynamical driving force, in what came to be known as a \emph{stress-assisted transformation}.

It has also been suggested that a second, distinct mechanism manifests in certain conditions (mainly depending on temperature): the \emph{strain-induced transformation} \citep{Olson1972}. Here, the transformation happens at stresses higher than the austenite yield stress, leading \citeauthor{Olson1972} to propose that slip activity in the austenite generates martensite nucleation sites as a result of shear band intersections. At lower temperatures, where transformation happens before yielding, stress-assisted transformations are dominant; conversely, at higher temperatures, strain-induced transformations are more significant \citep{Olson1978}. Some popular constitutive models derive from these observations \citep{Olson1975a,Stringfellow1992}.

Although the two types of transformation produce martensites with different morphologies and characteristics \citep{Maxwell1974,Snell1977}, there is no consensus in the literature on whether they are indeed due to two different mechanisms, or whether all mechanically induced transformations can be explained in terms of a single stress-based criterion \citep{Tamura1982,Chatterjee2007,Kundu2014}. Many authors proposed transformation models based on the latter assumption that positively reproduce experimental data for TRIP steels and meta-stable austenitic stainless steels \citep{Tamura1982,Chatterjee2007,Perdahcioglu2008,Geijselaers2009}.

In addition to those mentioned above, there is a large number of constitutive models for materials undergoing mechanically induced \emph{non\hyp{}thermoelastic} martensitic transformations, encompassing a broad variety of approaches, ranging from phenomenological models that postulate the existence of internal variables \citep{Leblond1989,Leblond1989a,Fischer1990,Bhattacharya1994,Govindjee2001,Kubler2011} to those of a micro-mechanical nature \citep{Marketz1995,Fischer1998,Reisner1998,Cherkaoui1998,Cherkaoui2000a,Cherkaoui2000,Ganghoffer1998,Idesman1999,Idesman2000,Levitas1998,Levitas1998a,Levitas2002,Lani2007}. Many of the more recent models use a computational homogenisation framework, such as those of  \cite{Iwamoto2004,Suiker2005,Turteltaub2005,Turteltaub2006,Tjahjanto2008,Hallberg2007,Kouznetsova2008,Sun2008,Lee2010,Yadegari2012} and \cite{Perdahcioglu2012}. 
Most of these, however, are formulated in a small strains context, an assumption of questionable validity due to large local transformation deformations, even though the resulting macroscopic deformation is generally small. Additionally, many models use averaging techniques assuming the collective behaviour of the microscopic transformation variants, ignoring some of the fine-scale interactive complexity between them. Other frequently used ingredients are pre-defined functional forms for the evolution of the martensite volume fraction, which involve parameters that bear limited physical significance and need to be calibrated from experiments. In this context, a transformation model relying primarily on readily measurable properties of the parent and product lattices (that is, those resulting from well-established frameworks such as the PTMC), avoiding the introduction of phenomenological elements, would be a valuable addition to this collection. The model analysed here has been initially formulated by \citet{Adziman2014} and \citet{deBortoli2017} and its assessment is the main object of this contribution.

This article is organised as follows: in \cref{sec:formulation}, an overview of the main experimental observations motivating the model are presented, along with the resulting formulation. In \cref{sec:algorithmics}, some notes are made about the model's algorithmic treatment in the context of an implicit finite element discretisation. In \cref{sec:results}, a meta-stable austenitic stainless steel for which experimental data is available is analysed, with particular attention dedicated to determining its transformation locus. Finally, \cref{sec:conclusions} presents a summary of the main achievements of this contribution, along with perspectives for future developments.

\subsection{Notation} For the most part, the notation used is standard in modern continuum mechanics, where summation (Einstein) convention is assumed unless otherwise noted. Vectors are denoted by italic bold-face lower-case letters such as $\vec{v}$, while their upper-case counterparts are used for second-order tensors such as $\tenII{T}$. Scalars and scalar-valued functions are usually written as light-face italic letters such as $s$. The scalar product of two vectors is denoted by $\vec{u} \cdot \vec{v} = u_i v_i$, whereas the double contraction of two second-order tensors is written as $\tenII{S} : \tenII{T} = S_{ij} T_{ij}$. The tensor product of vectors $\vec{u}$ and $\vec{v}$ is denoted by $\tenII{T} = \vec{u} \otimes \vec{v}$, with Cartesian components $T_{ij} = u_i v_j$. The second-order identity tensor is written as $\tenII{I} = \delta_{ij} \vec{e}_i \otimes \vec{e}_j$, where $\vec{e}_i$ are the orthonormal basis vectors of the Cartesian coordinate system. The deviatoric projection of a symmetric second-order tensor $\tenII{S}$ is given by $\dev[\tenII{S}] \equiv \tenII{S} - \frac{1}{3} \left( \tenII{S} : \tenII{I} \right) \tenII{I}$. The time derivative of a quantity $q$ is represented by $\dot{q}$. A dot over a line indicates the time derivative of the quantity under the line; for instance,
\begin{equation*}
  \dot{\overbar{\tenII{S} \tenII{T}}} = \parc{}{t} \left[ \tenII{S} \tenII{T} \right] .
\end{equation*}

\section{Formulation}\label{sec:formulation}

\subsection{Mechanically induced martensitic transformations}

This work focuses on modelling \emph{irreversible} stress-assisted/strain-induced austenite\hyp{}to\hyp{}martensite phase transformations that occur in ferrous alloys under \emph{isothermal mechanical loading}. This encompasses, for instance, both fully-austenitic and multi-phase steels partly composed of meta-stable austenite, such as transformation-induced plasticity (TRIP) steels. These ferrous alloys generally have face-centred cubic (FCC) austenite phases that transform to body-centred tetragonal (BCT) martensite. For these materials, the martensite tetragonality is directly correlated to the alloy's carbon content, so that in low carbon steels BCT martensite can be approximated as body-centred cubic (BCC) \citep{Nishiyama1978}.

\begin{figure} [h]
  \centering
  \includegraphics[width=0.85\textwidth]{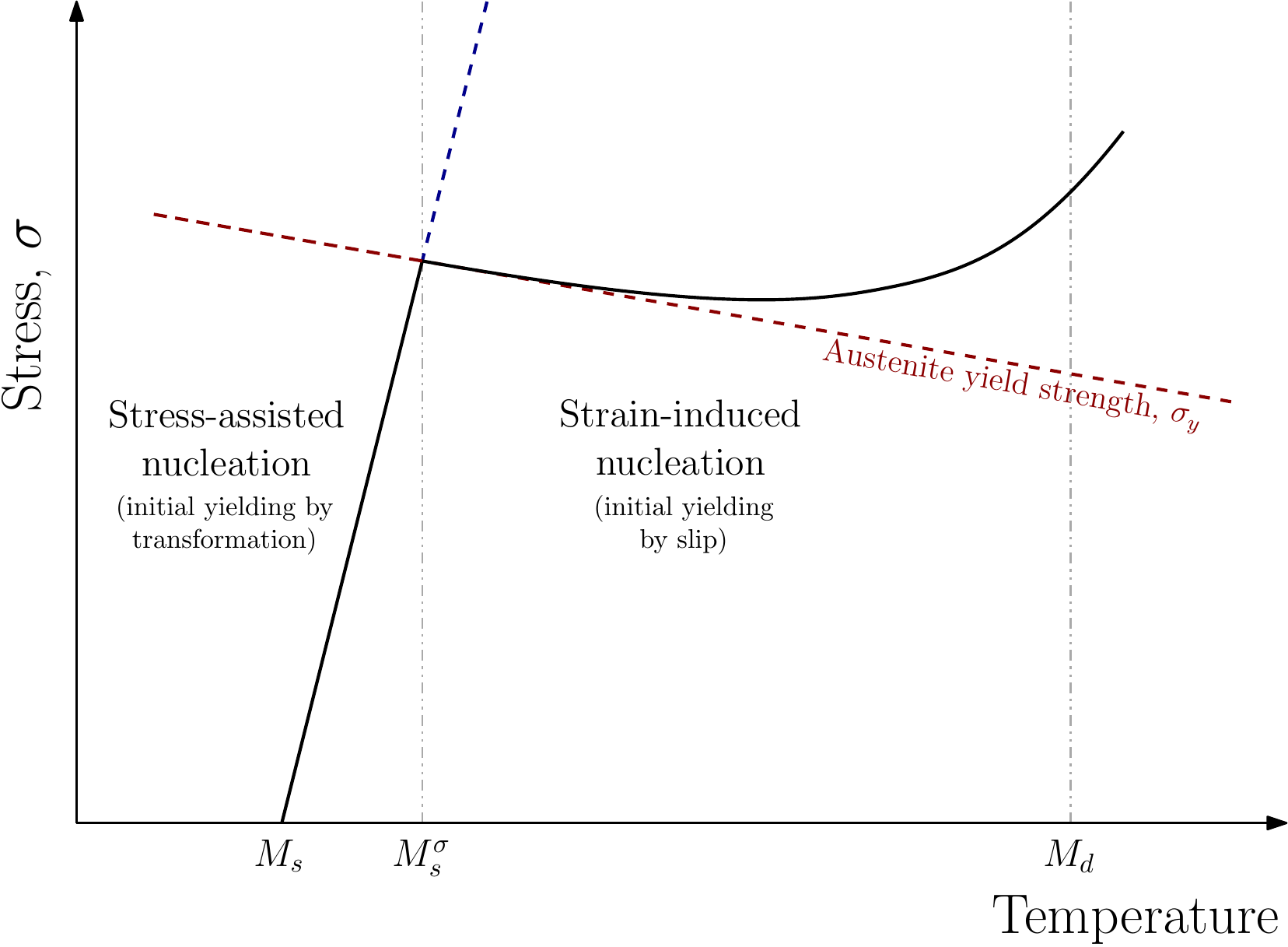}
  \caption{Schematic representation of stress-assisted and strain-induced martensitic transformation regimes, according to the relationship between martensite nucleation stress and the austenite yield stress, which varies with the temperature.} \label{fig:strain-assisted}
\end{figure}

\begin{figure} [h]
  \centering
  \includegraphics[width=0.85\textwidth]{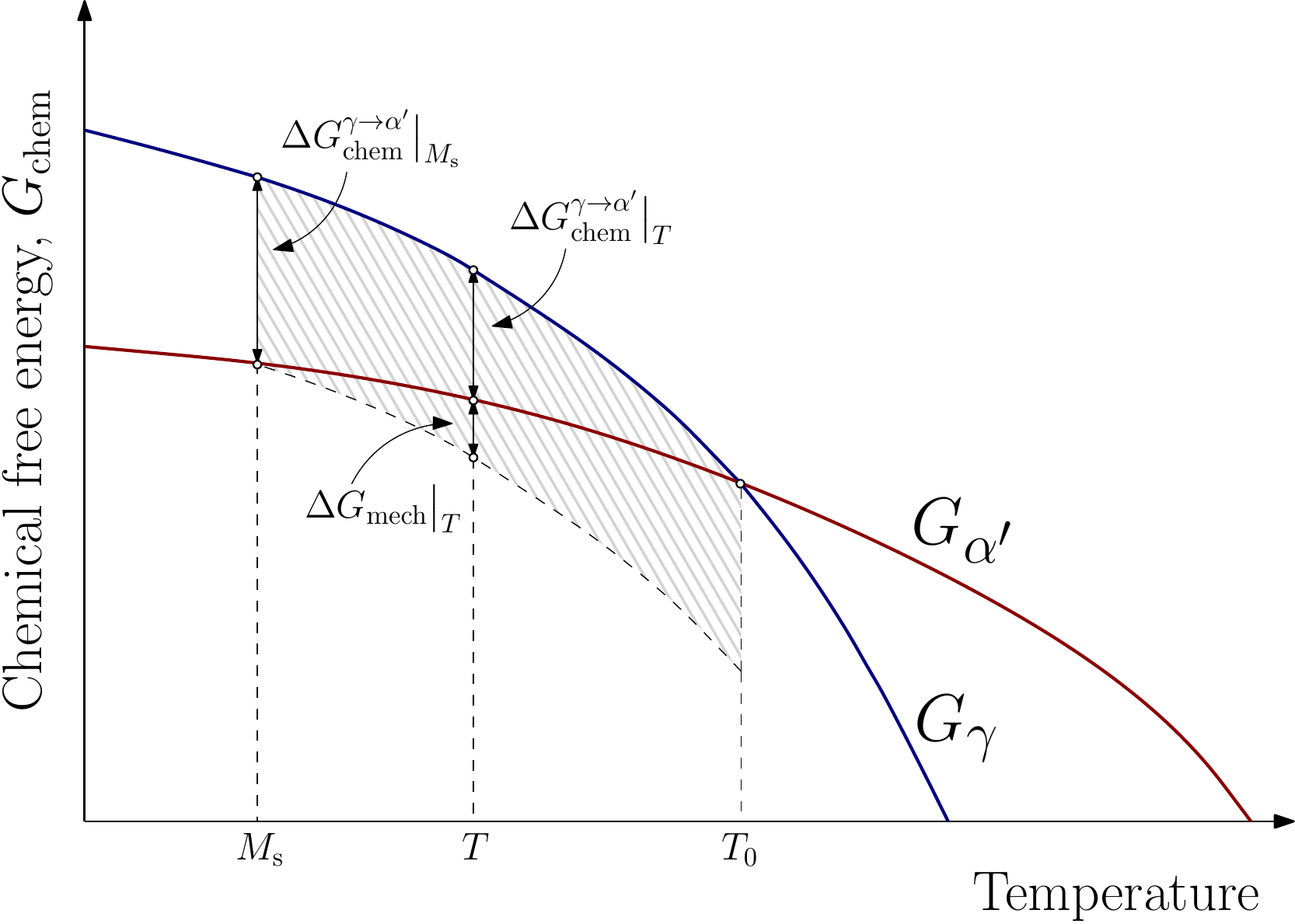}
  \caption{Chemical free-energy of austenite ($\gamma$) and martensite ($\alpha'$) phases as a function of temperature. Between temperatures $M_\text{s}$ and $T_0$, it is possible to induce transformation from $\gamma$ to $\alpha'$ with supplementary energy from external mechanical loadings ($\DGm$).} \label{fig:delta-gmech}
\end{figure}

The phenomena of importance in these transformations take place at the crystal lattice scale, where the movement of atoms dictates the relation between parent (austenite) and product (martensite) lattice properties. Using this lattice-scale information, a continuum model is constructed on the crystal scale. Its main driving assumptions are based on the following important experimental observations:
\begin{itemize}
  \item Plasticity can occur in the parent austenite phase before the martensitic transformation onset, depending on the alloy and on conditions such as testing temperature, as shown in \cref{fig:strain-assisted}. Below the temperature $M_\text{s}$, martensite forms with no external loading; through applied stress, it can be raised until the limit $M_\text{d}$, also known as the martensite deformation temperature. The previously described phenomena have been observed in TRIP steels \citep{Chatterjee2007} and metastable austenitic stainless steels \citep{Perdahcioglu2008a,Perdahcioglu2008}. It appears that modelling such plasticity is of critical importance to capture the complex interactions between crystallographic slip and the martensitic transformation. Additionally, modelling the inelastic behaviour of the austenite may shed some light on the question of whether stress-assisted and strain-induced martensitic transformations are indeed two distinct mechanisms, or they both can be explained in a unified formulation following an energy-based transformation criterion.
  \item The geometric relation between the austenite and martensite lattices can be satisfactorily predicted using the PTMC, resulting in a set of habit planes and transformation deformation vectors. These define the volumetric and shear strains involved in the transformation and their direction. The PTMC predicts that, for given parent and product lattices, there are multiple \emph{transformation systems}\footnote{In the materials science literature, the terms \emph{transformation system} and \emph{variant} are often used interchangeably, unlike their usage in group theory (as in, for instance, \cite{Hane1998}).} -- 24 in the case of an FCC-to-BCT/BCC transformation.
  \item A critical energy required to initiate the transformation is assumed to exist. At a given temperature, this energy barrier is considered a material property, depending on both parent and product phases relative chemical free-energies. This is illustrated in \cref{fig:delta-gmech}, where the chemical free-energy of the austenite ($\gamma$) and martensite ($\alpha'$) phases as a function of temperature is shown. The scalar $M_\text{s}$ is the temperature at which transformation starts spontaneously, while $T_0$ is the temperature at which both phases are in equilibrium. Above $T_0$, the austenite has lower chemical free energy than the martensite -- being therefore more stable -- so transformation is not possible. Between $M_\text{s}$ and $T_0$, the chemical free-energy difference between the phases is not enough for spontaneous transformation; external mechanical loads can, however, supply the remaining necessary energy. Once again, as clearly seen in the figure, the energy necessary to trigger a transformation is a function of the temperature: $\DGm \big|_T,\text{ } T \in [M_\text{s}, T_0]$. At a given temperature, $\DGm$ is equal to the stress power due to external loadings at which transformation starts. This is a widely accepted idea in the literature \citep{Bhadeshia2009,Tamura1982}, with many proposed models based on it (for instance, \cite{Kouznetsova2008,Perdahcioglu2012}).
  \item Given the existence of this critical transformation energy, a set of \emph{transformation functions} (cf. yield functions in the slip plasticity case) is postulated, generalising for arbitrary stress states what was initially proposed by \citet{Patel1953}. At a given material point, the selection of active set of transformation systems is also performed by using transformation functions: only the most favourably oriented variants with respect to the local stress state transform.
  \item At a given material point, once transformation starts in any system, plastic slip activity is considered to cease in the underlying crystal lattice; when the transformation finishes, the resulting fully martensitic material behaves elastically again.
\end{itemize}

\subsection{Kinematics}

Following standard practice in crystal plasticity modelling, the total deformation gradient $\tenII{F}$ at a given material point is multiplicatively decomposed to account for individual contributions of the deformation mechanisms under consideration\footnote{See \citet{Reina2014} for a rigorous micro-mechanical justification of the multiplicative split in the context of slip plasticity.}:
\begin{equation} \label{eq:F_decomposition}
  \tenII{F} = \Fe \Ftr \Fpa.
\end{equation}
Here, $\Fe$, $\Ftr$ and $\Fpa$ are, respectively, the elastic, martensitic transformation, and austenite plasticity deformation gradients.

\subsubsection{Austenite plastic slip}
In the austenite single crystal, plastic slip happens along preferred crystallographic directions. Each slip system $\alpha$ is defined by unit vectors $\ma$ and $\sa$ denoting, respectively, the normal vector to the slip plane and the slip direction. As the slip direction is contained in the slip plane, $\ma \cdot \sa = 0$ meaning these deformations are \emph{isochoric}. The usual flow rule from crystal plasticity is assumed \citep{Asaro1985}:
\begin{equation} \label{eq:plastic_flow_rule}
  \FpaDot = \left[ \sum_{\alpha=1}^{\nss} \gm{pr} \sma \right] \Fpa,
\end{equation}
where $\nss$ is the number of slip systems in the crystal under consideration\footnote{It is implied that the summation above is done only on \emph{active} slip systems (i.e. those where there currently is slip activity), as otherwise $\gm{pr} = 0$.}.

In the previous equation, the plastic shear rates $\gm{pr}$ satisfy the complementarity conditions\footnote{Summation of the repeated index is not implied in the last equation.}:
\begin{equation} \label{eq:yield_function}
  \yf{p} \leq 0, \qquad
  \gm{pr} \geq 0, \qquad
  \yf{p} \gm{pr} = 0, \qquad i = 1, \dots, \nss,
\end{equation}
where $\yf{p}$ denotes the slip system's yield function. Since modelling plastic deformations in FCC austenite phases is of primary interest in this study, crystallographic slip is assumed to follow Schmid's law\footnote{Schmid's law's applicability to FCC crystals is well established, in contrast to the BCC case, where \emph{non-Schmid effects} are often important \citep{Dao1993}. Modified versions of Schmid's law have been proposed for this purpose in the literature (for example, see \cite{Yalcinkaya2008}), but are beyond the scope of this work.}:
\begin{equation}
  \yf{p} \left( \tau^\alpha, \tau_y^\alpha \right) \equiv |\tau^\alpha| - \tau_y^\alpha,
\end{equation}
with the resolved Schmid shear stress $\tau^\alpha$ defined as:
\begin{equation}
  \tau^{\alpha} \equiv \left( \tenII{\sigma} \ma \right) \cdot \sa = \tenII{\sigma} : \left( \sma \right),
\end{equation}
where $\tenII{\sigma}$ is the Cauchy stress tensor. For simplicity, it is assumed that the critical resolved shear stress $\tau_y^\alpha$ in \cref{eq:yield_function} follows an isotropic hardening of Taylor type \citep{Taylor1938}, meaning the critical resolved shear stress $\tau_y^{\alpha}$ has the same value for all systems,
\begin{equation}
  \tau_y^{\alpha} = \tau_y (\gamma_\text{pa}),
\end{equation}
and is a function of the accumulated plastic slip $\gamma_\text{pa}$:
\begin{equation}
  \gamma_\text{pa} \equiv \int_0^t \sum_{\alpha=1}^{\nss} |\gm{pr}| \diff t.
\end{equation}

\subsubsection{Elastic behaviour}

The reversible part of the crystal deformation is modelled using a regularised neo-Hookean hyperelastic potential, resulting in the following expression for the Kirchhoff stress $\tenII{\tau}$:
\begin{equation} \label{eq:neo-hookean}
  \tenII{\tau} = G \dev{[ \Beiso ]} + K (\ln J^\text{e}) \tenII{I},
\end{equation}
where $G$ and $K$ are the shear and bulk moduli, $\,J^\text{e} \equiv \det[\Fe]$, $\,\Beiso \equiv \Feiso \FeTiso$, and $\,\Feiso \equiv \left(J^\text{e }\right)^{-\frac{1}{3}} \Fe$. Although the lattice elastic rotations are generally large, its elastic distortions are usually negligible \citep{Kim2014}. As such, this choice of potential is primarily motivated by convenience, as it entails a simple expression for the resolved Schmid stresses:
\begin{equation}
  \tau^{\alpha} = G \left( \Feiso \sa \cdot \Feiso \ma \right).
\end{equation}

\subsubsection{Martensitic transformation kinematics}

The crystallographic theory of martensite predicts, for given parent and product lattices, a set of $\nts$ transformation systems; these are characterised by a habit plane with unit normal vector $\mi$, along with shear directions $\ssi$, for $i = 1, 2, \dots, \nts$. The number of transformation systems $\nts$ depends on the respective symmetries of the parent and product lattices, while the system vectors depend on the relation between their lattice parameters \citep{Wechsler1953,Ball1987}.

\begin{figure} [h]
  \centering
  \includegraphics[width=0.95\textwidth]{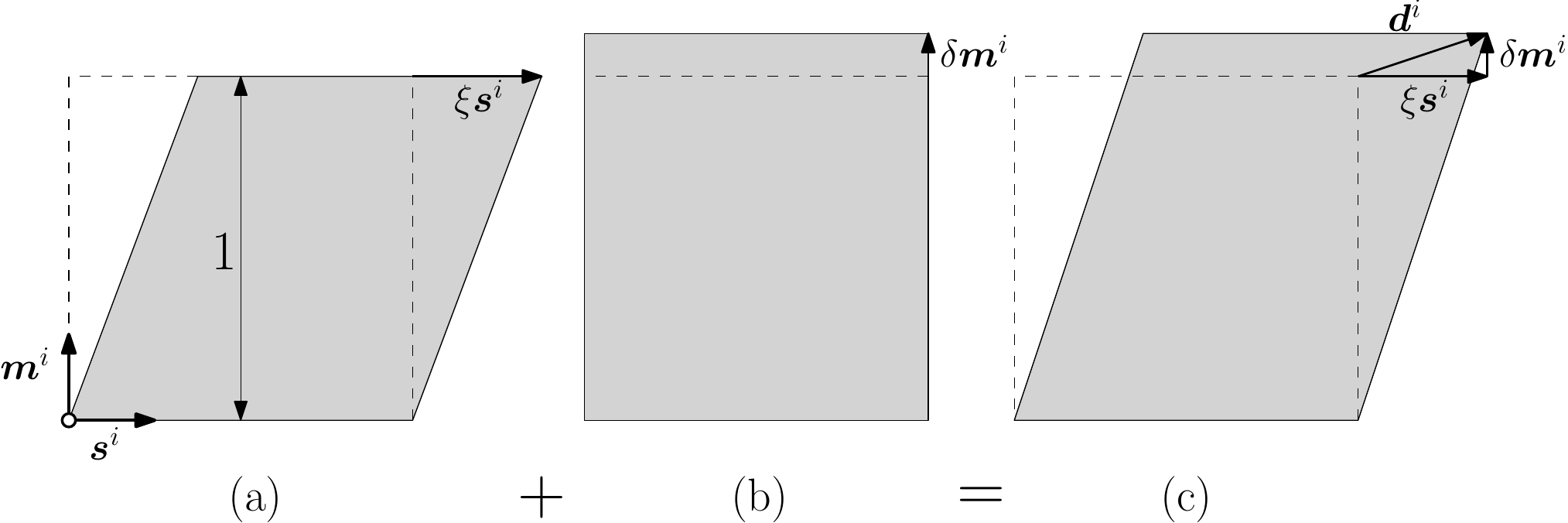}
  \caption{Decomposition of the total transformation deformation gradient (c) for a given system into: (a) a shear in the habit plane, in the direction $\ssi$, and (b) a volumetric deformation, normal to the habit plane $\mi$.} \label{fig:ftr_decomp}
\end{figure}

The deformation due to the martensitic transformation on a system $i$, depicted in \cref{fig:ftr_decomp}, consists of a habit plane shear of magnitude $\xi$ in the direction $\ssi$, along with an expansion of magnitude $\delta$ in the direction $\mi$:
\begin{equation} \label{eq:ftr-def}
  \Ftr = \tenII{I} + \xi \ssi \otimes \mi +  \delta \mi \otimes \mi = \tenII{I} + \dmi,
\end{equation}
where the transformation deformation vector $\di \equiv \xi \ssi + \delta \mi$ is defined. Unlike the case of \cref{eq:plastic_flow_rule}, vectors $\di$ and $\mi$ are \emph{not orthogonal}, due to the transformation's volumetric component. In fact, expressing $\Ftr$ in the orthonormal basis $\{\ssi, \mi, \ssi \times \mi\}$, it is evident that the transformation deformation gradient is \emph{not isochoric}:
\begin{equation}
  \Ftr = \begin{bmatrix}
    1 &  \xi       & 0 \\
    0 & 1 + \delta & 0 \\
    0 &   0        & 1
  \end{bmatrix} \quad \Longrightarrow \quad \det[\Ftr] = 1 + \delta.
\end{equation}

\subsubsection{Transformation multipliers}
At a given material point, the transformation is considered to potentially happen in multiple systems concurrently; its evolution can thus be modelled by introducing \emph{transformation multipliers} $\gm{t} \in [0,1]$, leading to \cref{eq:ftr-def} being rewritten as:
\begin{equation} \label{eq:F_tr}
  \Ftr = \tenII{I} + \sum_{i=1}^{\nts}\gm{t} \dmi.
\end{equation}
These multipliers can be interpreted as local volume fractions of martensite transformed along each system. The total martensite and austenite volume fractions are then given, respectively, by:
\begin{equation}
  \gamma_\text{m} = \sum_{i=1}^{\nts}\gm{t}, \qquad \gamma_\text{a} = 1 - \gamma_\text{m},
\end{equation}
which follow the evident restrictions:
\begin{equation}
  0 \leq \gm{t} \leq 1, \qquad 0 \leq \gamma_\text{m} \leq 1, \qquad 0 \leq \gamma_\text{a} \leq 1.
\end{equation}
Note that the transformation irreversibility is reflected by the fact that its multipliers are \emph{monotonically increasing} functions of time:
\begin{equation}
  \gm{tr} \geq 0, \quad \forall \, i \in 1, \dots, \nts.
\end{equation}

\subsection{Thermodynamical considerations}

\subsubsection{Free-energy function}
The mechanical free-energy density $\psi$ is postulated as independent from both crystallographic slip and martensitic transformation status, being thus a function of the elastic deformation gradient only: $\psi = \psi \left( \Fe \right)$. Using the multiplicative decomposition from \cref{eq:F_decomposition}, its rate of change during the martensitic transformation evolution is given by:
\begin{align}
  \dot{\psi} & = \parc{\psi}{\Fe} : \FeDot
               = \parc{\psi}{\Fe} :
                 \left( \dot{\tenII{F}} \Fpam \Ftrm +  \tenII{F} \Fpam \dot{ \overbar{ \Ftrm } } \right) \notag \\
             & = \parc{\psi}{\Fe} \FtrmT \FpamT : \dot{\tenII{F}}
                 - \FeT \parc{\psi}{\Fe} \FtrmT : \FtrDot,
\end{align}
where it is assumed that \emph{plastic slip ceases as soon as the transformation starts in any system} (so $\FpaDot = \tenII{0}$).

\subsubsection{Dissipation rate}
Combining the above with the Clausius--Duhem inequality yields:
\begin{equation}
  \dot{D} = \left( \tenII{P} - \bar{\rho} \parc{\psi}{\Fe} \FtrmT \FpamT \right) : \dot{\tenII{F}}
            + \FeT \bar{\rho} \parc{\psi}{\Fe} \FtrmT : \FtrDot \geq 0,
\end{equation}
which must hold for every admissible process and therefore for arbitrary $\dot{\tenII{F}}$ and $\FtrDot$. On top of the usual constitutive equation for the first Piola--Kirchhoff stress $\tenII{P}$, a simplified expression for the dissipation rate $\dot{D}$ ensues:
\begin{equation} \label{eq:dissipation-rate}
  \dot{D} = \FeT \bar{\rho} \parc{\psi}{\Fe} \FtrmT : \FtrDot = \tenII{T} : \FtrDot \geq 0,
\end{equation}
from which $\tenII{T}$, the \emph{work conjugate stress of the transformation deformation gradient}, can be defined:
\begin{equation}
  \tenII{T} \equiv \FeT \tenII{P} \FpaT = \FeT \tenII{\tau} \FmT \FpaT,
\end{equation}
where $\tenII{\tau}$ is the Kirchhoff stress tensor.

\subsubsection{Transformation functions; mechanical dissipation consistency}
As is clear from \cref{fig:delta-gmech}, for a given temperature between $M_\text{s}$ and $T_0$, the activation of the martensitic transformation requires a critical mechanical energy $\DGm$ (per unit reference volume). A constitutive model that dissipates precisely this amount of energy over the course of the martensitic transformation can be devised by postulating the existence of $\nts$ \emph{transformation functions} (cf. yield functions in plasticity) of the type:
\begin{equation} \label{eq:transformation-function}
  \yf{t}(\tenII{T}) \equiv \tenII{T} : \left(\dmi \right) - \DGm = T^i - \DGm,
\end{equation}
together with the usual loading/unloading conditions\footnote{Once more, summation of the repeated index is not implied in the last equation.}:
\begin{equation}
  \yf{t} \leq 0, \qquad
  \gm{tr} \geq 0, \qquad
  \yf{t} \gm{tr} = 0, \qquad i = 1, \dots, \nts.
\end{equation}
In the above, $T^i$ is the projection of the transformation conjugate stress tensor into the transformation system $i$: $\,T^i \equiv \tenII{T} : \left(\dmi \right)$. These projections can be interpreted as the martensitic transformation analogous counterparts to the resolved Schmid stresses in slip plasticity.

The transformation flow rule follows from \cref{eq:F_tr}:
\begin{equation}
  \FtrDot = \sum_{i=1}^{\nts} \gm{tr} \dmi
             = \sum_{i=1}^{\nts} \gm{tr} \parc{\yf{t}}{\tenII{T}},
\end{equation}
where it is considered that $\di$ and $\mi$ are constant during the transformation. From the last equality above, it follows that the transformation flow rule is \emph{associative}.

The consistency of the dissipation follows from \cref{eq:dissipation-rate,eq:transformation-function}:
\begin{equation}
  \dot{D} = \tenII{T} : \FtrDot = \sum_{i=1}^{\nts} \gm{tr} T^i
          = \sum_{i=1}^{\nts} \gm{tr} \DGm.
\end{equation}
Integrating this dissipation rate through the course of the transformation, starting from a fully austenitic state ($\gamma_\text{m} = 0$) and ending with pure martensite ($\gamma_\text{m} = 1$), results in:
\begin{align}
  D \equiv \int_{t_0}^{t_f} \dot{D}(t) \diff t 
    & = \DGm \int_{t_0}^{t_f} \sum_{i=1}^{\nts} \gm{tr} \diff t
      = \DGm \int_{t_0}^{t_f} \dot{\gamma}_\text{m} \diff t \notag \\
    & = \DGm \left( \gamma_\text{m} \bigg\vert_{t = t_f} - \gamma_\text{m} \bigg\vert_{t = t_0} \right) = \DGm.
\end{align}

It is therefore evident that the total mechanical energy dissipated during transformation is rigorously equal to the energy density required for transformation, $\DGm$.

\subsubsection{Relation to the Mohr--Coulomb yield criterion}
As will be clear momentarily, the above transformation functions are connected to the classical Mohr--Coulomb yield function
\begin{equation} \label{eq:MC-yield}
  \Phi^{MC} = \tau + \sigma_n \tan \phi - c,
\end{equation}
where $\tau$ and $\sigma_n$ are the shear and normal (assumed tensile positive) stresses at any given plane through a material point. Material parameter $c$ represents the medium's cohesion, whereas $\phi$ is known as its frictional angle.

Consider now the martensitic transformation functions \labelcref{eq:transformation-function}. Expressing the stress tensor $\tenII{T}$ on the orthonormal basis associated to the habit plane of transformation system $i$, $\{\ssi, \mi, \ti\}$, where $\ti = \ssi \times \mi$, yields:
\begin{align}
  \yf{t}(T^i) & =
  \begin{bmatrix}
    T_{s^i s^i }  &  T_{s^i m^i }  &  T_{s^i t^i }  \\
    T_{m^i s^i }  &  T_{m^i m^i }  &  T_{m^i t^i }  \\
    T_{t^i s^i }  &  T_{t^i m^i }  &  T_{t^i t^i }
  \end{bmatrix} : \begin{bmatrix} 
    0  &  \xi    &  0  \\
    0  &  \delta &  0  \\
    0  &  0      &  0
  \end{bmatrix} - \DGm \notag \\
  & = \xi T_{s^i m^i }  + \delta T_{m^i m^i }  - \DGm. \label{eq:MC-comparison}
\end{align}

Clearly, $T_{s^i m^i }$ and $T_{m^i m^i }$ are shear and normal internal tractions acting in the habit plane, respectively. The transformation starts when the work due to these tractions reaches the energy barrier $\DGm$. For the Mohr--Coulomb criterion, yield starts when the critical combination \eqref{eq:MC-yield} is reached on \emph{any plane} at a material point, whereas for transformation criterion the combination must occur on a \emph{transformation habit plane}. For a material with randomly oriented austenite crystals, the transformation surface can thus be expected to converge to a Mohr--Coulomb-type locus with an increase in the number of grains, as any orientation can be made arbitrarily close to a habit plane.

Additionally, comparing \cref{eq:MC-yield,eq:MC-comparison}, it is apparent that the parameters of the equivalent Mohr--Coulomb yield surface are:
\begin{equation} \label{eq:MC-params}
  \phi = \tan^{-1}\left( \frac{\delta}{\xi}\right), \quad c = \frac{\DGm}{\xi}.
\end{equation}
Thus, $\phi$ is the angle between the transformation shear direction $\ssi $ and its total deformation vector $ \di  \equiv \xi \ssi  + \delta \mi $. The cohesion parameter $c$ can be interpreted, analogously to the Mohr--Coulomb model, as the habit plane shear stress that is necessary for transformation to start in the absence of normal stresses to the habit plane (as can be seen by setting $T_{m^i m^i } = 0$ in \cref{eq:MC-comparison}).

It is interesting to note that this kind of transformation surface has been reported in shape\hyp{}memory alloys undergoing mechanically induced martensitic transformations (for instance, see \cite{Orgeas1998}). The materials tested by \citeauthor{Orgeas1998} display almost no pressure-sensitivity in their transformation surfaces. This can be understood from the fact that the martensitic transformations in shape\hyp{}memory alloys involve practically no dilatation ($\delta \approx 0$). Thus, $\phi \approx \SI{0}{\degree}$ and the resulting Mohr--Coulomb surface is weakly pressure-dependent.

\section{Algorithmic treatment} \label{sec:algorithmics}

The constitutive equations above are integrated using a fully implicit backward-Euler method. For the austenite plasticity, an integrator based on the exponential return-mapping is used as it is exactly volume-preserving, satisfying the plastic deformation incompressibility for any step size \citep{Miehe1996b}. Full details of the associated linearised equations and algorithmic procedures are presented in \citet{deBortoli2017} and \citet{Adziman2014}.

In the above rate-independent model, the set of active systems (i.e. systems for which slip/transformation is currently active) needs to be known for the stress update to be performed. This problem has no trivial solution, due to the high number of potentially active systems\footnote{For instance, the 12 slip systems in FCC crystals and 24 martensitic transformation variants in the FCC-to-BCC case.}. Namely, given the necessary linear dependencies between the multiple slip/transformation systems, there are in general various combinations of multipliers that produce the same incremental plastic/transformation deformation gradient and resulting stress tensor. Additionally, Newton-type algorithms have convergence difficulties when the active set determination is done concurrently with the regular iterations.
Algorithms that perform the search of active systems are still an open area of investigation in crystal plasticity, having been extensively studied primarily in the context of slip deformation \citep{Borja1993,Anand1996,Akpama2016}.

A common way to avoid the above issues relies on visco-plastic regularisations of the original problem \citep{Asaro1985}. Thus, a rate-dependent formulation is used, even though the materials of interest in this work are not modelled in conditions where actual rate-dependent phenomena (like creep) are of importance. The rate-independent formulation can be recovered in the limit of vanishing rate-sensitivity parameters. This approach has the disadvantage, however, of leading to numerical difficulties due to the resulting stiffness of the system of return-mapping equations \citep{Steinmann1996}.

To that end, a slip-rate law proposed by \cite{Peric1993} is adopted for both the evolution of slip plastic multipliers:
\begin{equation}
  \gm{pr} =
  \begin{cases} \displaystyle
    \frac{1}{\mu_\text{pa}} \left[  \left( \frac{ |\tau^{\alpha}|}{\tau_y}\right)^{1/\epsilon_\text{pa}} -1 \right] \sign[\tau^{\alpha}]
      & \text{if } \yf{p} \left(\tau^{\alpha}, \tau_y \right) \geq 0 \\
    0
      & \text{if } \yf{p} \left(\tau^{\alpha}, \tau_y \right) < 0
  \end{cases}, \label{eq:gammadot-pa}
\end{equation}
and that of the transformation multipliers:
\begin{equation}
  \gm{tr} =
  \begin{cases} \displaystyle
    \frac{1}{\mu_\text{tr}} \left[  \left( \frac{ T^i }{\DGm}\right)^{1/\epsilon_\text{tr}} -1 \right]
      & \text{if } \yf{t} \left( T^i \right) \geq 0 \\
    0
      & \text{if } \yf{t} \left( T^i \right) < 0
  \end{cases}. \label{eq:gammadot-tr}
\end{equation}
Material parameters $\mu_\text{pa}$ and $\mu_\text{tr}$ are viscosity-related (with time dimensions), while $\epsilon_\text{pa}$ and $\epsilon_\text{tr}$ are non-dimensional rate-sensitivity parameters. The rate-independent limit of the above expressions is reached as either $\mu_\text{pa} \to 0$, $\mu_\text{tr} \to 0$ or $\epsilon_\text{pa} \to 0$, $\epsilon_\text{tr} \to 0$. Here, the proper choice of these parameters is a trade-off between obtaining sufficiently rate-independent solutions and computational efficiency. Note that the presence of the absolute value and $\sign$ functions in \cref{eq:gammadot-pa} is due to the fact that crystallographic slip can happen in either direction along a particular slip system. The same, however, does not apply to the martensitic transformation, leading to the differences between \cref{eq:gammadot-pa,eq:gammadot-tr}.

Another noteworthy algorithmic aspect is the assumption governing the coupling of transformation and plasticity. Namely, once the transformation criterion is satisfied at a given material point, the austenite plasticity in it ceases and only the transformation can progress. This simplifies considerably the model's computational treatment, as the coupling between the two phenomena does not have to be dealt with explicitly in the stress integration procedure. Nevertheless, this assumption might lead to problems due to the sudden change in the material's stiffness, especially if it has already undergone a large amount of plastic slip. In a polycrystalline material, this effect's importance is expected to be diluted in the homogenised material response.

\section{Results} \label{sec:results}

The martensitic transformation constitutive model is now applied to the analysis of an ASTM A-564 alloy (also known as Sandvik Nanoflex), a meta-stable austenitic stainless steel for which \citeauthor{Perdahcioglu2008} published extensive experimental data \citep{Perdahcioglu2008a,Perdahcioglu2008,Geijselaers2009,Perdahcioglu2012}. Its chemical composition is given in \cref{table:sandvik-nanoflex-composition}.

\begin{table} [h]
  \caption{Chemical composition of the meta-stable austenitic stainless steel ASTM A-564, in weight \%.}
  \centering
  \begin{tabular}[c]{c c c c c c c c}
    \rule{0pt}{0.8\normalbaselineskip}
    C+N  &  Cr  &  Ni  &  Mo  &  Cu  &  Ti  &  Al  &  Si  \\
    \hline
    \rule{0pt}{0.8\normalbaselineskip}
    <0.05  & 12.0 &  9.1 &  4.0 &  2.0 &  0.9 &  0.4 &  <0.5  \\
  \end{tabular}
  \label{table:sandvik-nanoflex-composition}
\end{table}

Their experiments include multiple tests ranging from simple shear to plane strain tension, showing clear evidence that this alloy undergoes mechanically induced martensitic transformations \citep{Geijselaers2009}. Additionally, this material displays considerable plastic deformation before the martensitic transformation onset, at equivalent strains of the order of 5\%, mostly due to slip \citep{Perdahcioglu2012}. Under these conditions, it provides an opportunity to test how a stress-assisted transformation criterion behaves in the strain-induced regime. Of particular interest is the fact that \citeauthor{Perdahcioglu2008} found that the volume of martensite could be broadly considered a function of the mechanical driving force only.

For this alloy, the parent phase has an FCC lattice and thus the usual octahedral slip systems are considered (see \cref{table:octahedral_systems}, \cref{sec:appendix-systems}). The product phase can be considered approximately BCC, due to its low carbon content. The lattice parameters for the phases are:
\begin{equation}
  a_\gamma = \SI{3.59690}{\angstrom}, \quad a_{\alpha'} = \SI{2.87351}{\angstrom},
\end{equation}
from which the PTMC predicts the following families of habit plane and shape deformation vectors:
\begin{equation}
  \vec{m} = \{0.608, -0.178, 0.774\}, \quad \vec{d} = \{-0.156, 0.046, 0.159\}.
\end{equation}
The 24 transformation variants corresponding to these families are shown in \cref{table:transformation-systems} (\cref{sec:appendix-systems}). The transformation dilatation and shear -- equal for all variants -- is given by:
\begin{equation} \label{eq:nanoflex-params}
  \delta = \vec{d} \cdot \vec{m} \approx 0.02, \quad
  \xi = \sqrt{\vec{d} \cdot \vec{d} - \delta^2} \approx 0.226.
\end{equation}
It is worth noting the strain magnitude: the shear component, in particular, is very significant, putting into question the small strains assumption used by many constitutive models found in the literature.

\subsection{Mesh and material parameters}

\begin{figure} [h]
  \centering
  \begin{subfigure}{0.9\textwidth}
    \centering
    \includegraphics[width=\textwidth]{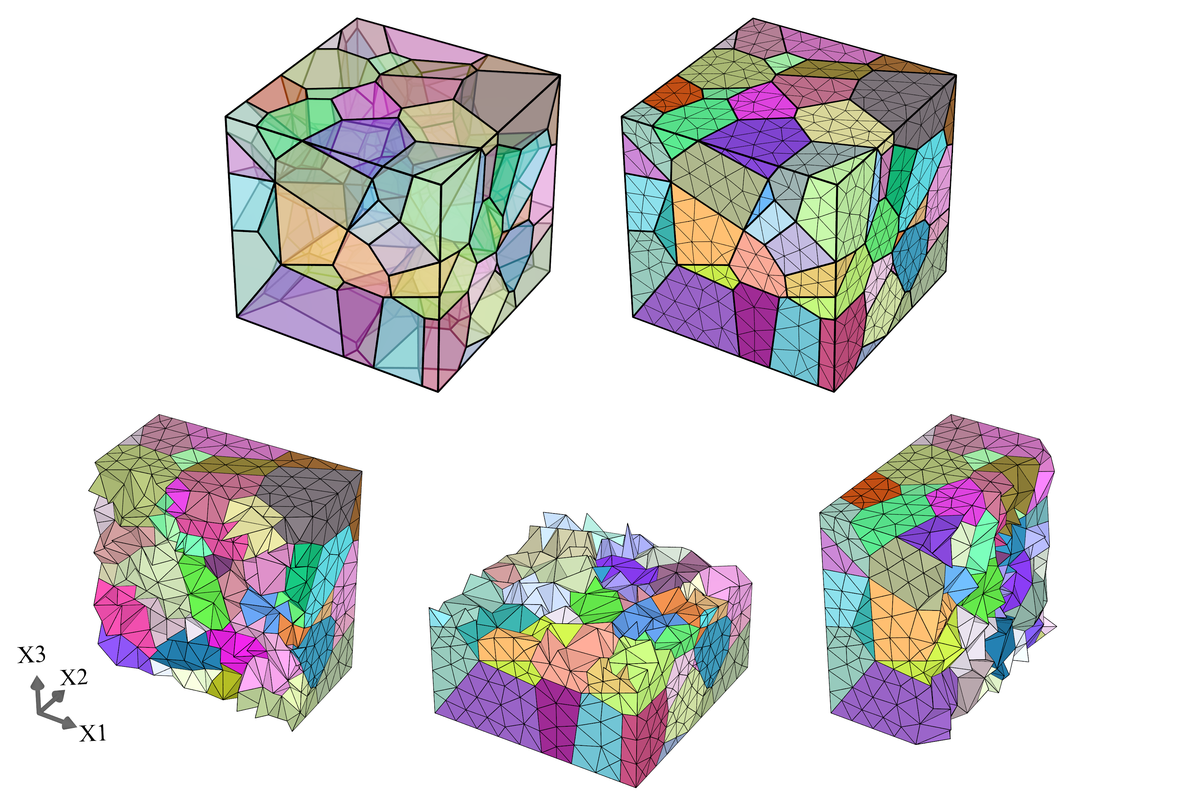}
    \caption{RVE 1 mesh: 100 grains, 13604 nodes and 9054 elements.}
  \end{subfigure}
  \par\bigskip
  \begin{subfigure}{0.9\textwidth}
    \centering
    \includegraphics[width=\textwidth]{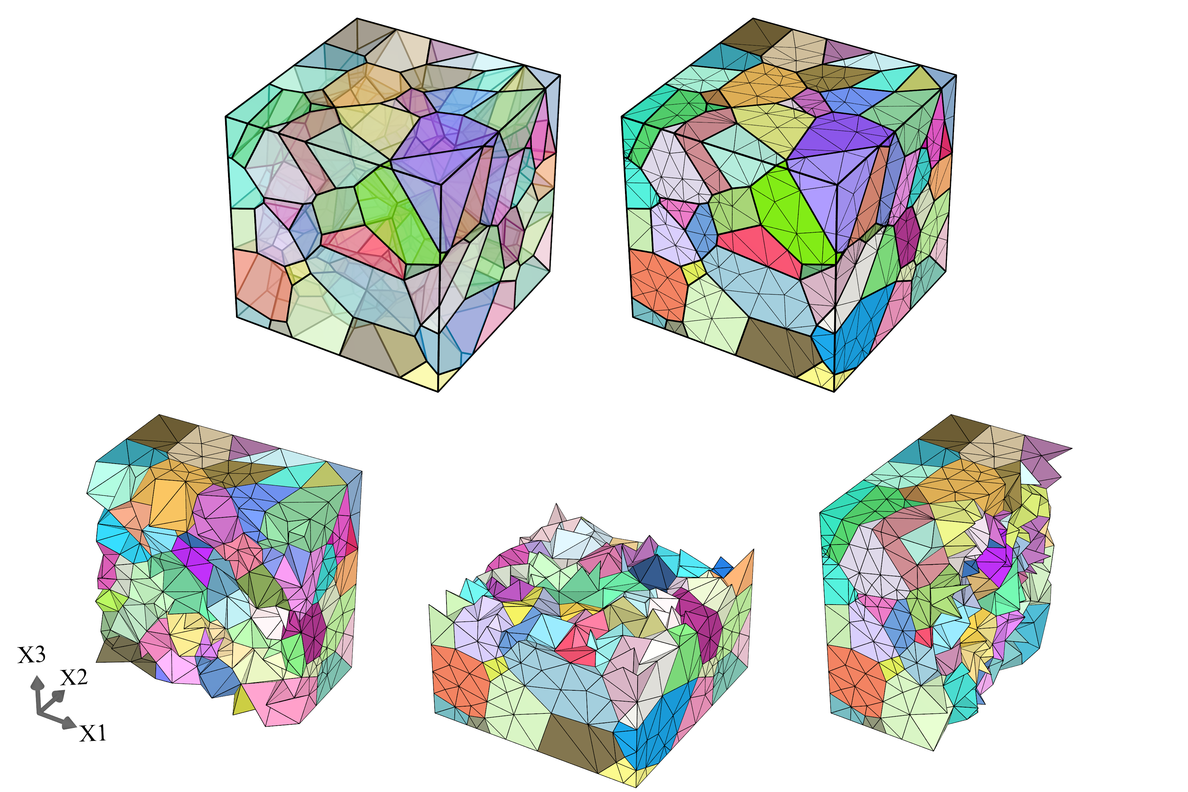}
    \caption{RVE 2 mesh: 150 grains, 13658 nodes and 9326 elements.}
  \end{subfigure}
  \caption{RVEs used in the analyses. Grain geometry and 10-node quadratic tetrahedral mesh are shown, along with cuts displaying internal element distribution. Grains are coloured according to crystallographic orientations: RGB colour values are related to the $x$, $y$ and $z$ components of their Rodrigues' vector, respectively.}
  \label{fig:mesh}
\end{figure}

To analyse this material in a computational homogenisation finite element code, representative volume elements (RVEs) are instantiated using \texttt{Neper}, a polycrystal generation and meshing library \citep{Quey2011}. As this material does not possess a particular texture, the polycrystal grains are chosen to be randomly oriented in the FCC cubic symmetry group. Due to the incompressibility of plastic slip, 10-node quadratic tetrahedral elements with 4 integration points are used to minimise spurious volumetric locking issues in the finite element formulation. To assess the appropriate number of grains necessary to capture the isotropic limit of the material behaviour, two RVEs are created consisting of 100 and 150 crystals using meshes with approximately the same numbers of elements and nodes. The polycrystal geometries and associated meshes are shown in \cref{fig:mesh}.

\FloatBarrier

In addition to the aforementioned crystallographic directions for both slip and martensitic transformation, elastic and transformation material parameters are taken from \cite{Perdahcioglu2012} and given in \cref{table:material-parameters}. The same authors also report stress-strain curves for a stable austenite phase in a simple shear test; this data is used to calibrate the hardening behaviour of the austenitic phase, assumed to follow a Nadai-Ludwik law:
\begin{equation} \label{eq:nadai-ludwik}
  \tau_y(\gamma_\text{pa}) = \tau_{y0} + K (\gamma_{\text{pa}0} + \gamma_\text{pa})^m.
\end{equation}
\citeauthor{Perdahcioglu2012} also calibrated one such hardening curve to this material. Their constitutive model for the austenite is, however, an isotropic plasticity model whose hardening curve is given in terms of effective Cauchy stress. As such, their reported parameters are of no immediate use to the calibration of a critical Schmid stress hardening curve, as the relation between both is not trivial. With that in mind, a new calibration is performed with the martensitic transformation model. To do that, an artificially high critical energy value is temporarily set so that the transformation criterion is not satisfied and only austenite plasticity persists. A simple inverse iterative procedure is then followed until a satisfactory fit is achieved, resulting in the parameters displayed in \cref{table:material-parameters}. It is important to note once again that, as this alloy's plasticity is of a rate-independent nature, the choice of the viscoplastic parameters in \cref{table:material-parameters} is motivated by a simple accuracy-computational time trade-off: the values reported here are found to make the response sufficiently close to rate-independent while maintaining the problem computationally tractable in terms of solution time.

\begin{table}
  \caption{Parameters used in the analyses.}
  \centering
  \begin{tabular}[c]{l l l c}
    \multicolumn{2}{c}{Material property}       & Value & Equation\\
    \hline\rule{0pt}{0.8\normalbaselineskip}
    \multirow{2}{*}{Elastic constants}          & $E$                   & \SI{210}{\giga\pascal} & \multirow{2}{*}{\eqref{eq:neo-hookean}}  \\
                                                & $\nu$                 & 0.3                    &                                          \\
    \hline\rule{0pt}{0.8\normalbaselineskip}
    \multirow{6}{*}{Austenite plasticity}       & $\tau_{y0}$           & \SI{100}{\mega\pascal} & \multirow{4}{*}{\eqref{eq:nadai-ludwik}} \\
                                                & $K$                   & \SI{195}{\mega\pascal} &                                          \\
                                                & $\gamma_{\text{pa}0}$ & 0.01                   &                                          \\
                                                & $m$                   & 0.6                    &                                          \\
                                                \cline{2-4}\rule{0pt}{0.8\normalbaselineskip}
                                                & $\mu_\text{pa}$       & \SI{0.2}{\second}      & \multirow{2}{*}{\eqref{eq:gammadot-pa}}  \\
                                                & $\epsilon_\text{pa}$  & 0.2                    &                                          \\
    \hline\rule{0pt}{0.8\normalbaselineskip}
    \multirow{3}{*}{Martensitic transformation} & $\DGm$                & \SI{56}{\mega\pascal}  & \eqref{eq:transformation-function}       \\
                                                \cline{2-4}\rule{0pt}{0.8\normalbaselineskip}
                                                & $\mu_\text{tr}$       & \SI{0.2}{\second}      & \multirow{2}{*}{\eqref{eq:gammadot-tr}}  \\
                                                & $\epsilon_\text{tr}$  & 0.2                    &                                          \\
  \end{tabular}
  \label{table:material-parameters}
\end{table}

In the numerical solution of this problem, the prescribed deformation gradient is a 50\% simple shear, resulting in an equivalent strain $\varepsilon_\text{eq} \approx 0.28$, where
\begin{equation}
  \varepsilon_\text{eq} \equiv \sqrt{\frac{2}{3} \dev \left[ \tenII{\varepsilon} \right] : \dev \left[ \tenII{\varepsilon} \right]},
\end{equation}
and $\tenII{\varepsilon}$ is the eulerian Hencky strain tensor:
\begin{equation}
  \tenII{\varepsilon} \equiv \frac{1}{2} \ln \left[ \tenII{F} \FT \right].
\end{equation}
This deformation is applied in $\SI{1}{\second}$, resulting in an average strain rate of approximately $\SI{0.28}{\per\second}$ in the RVE.

This simple shear test is analysed using different homogenisation boundary conditions: uniform strain (Taylor), linear boundary displacements, periodic boundary displacement fluctuations, and uniform boundary tractions \citep{deSouzaNeto2010}. As the polycrystal meshes do not have one-to-one nodal correspondence between opposing RVE faces, the periodic boundary condition is not trivially enforced. This is nevertheless the boundary condition of most interest, as it converges faster than the others to the representative properties of the homogenised medium \citep{Terada2000}, so a mortar-based formulation proposed by \citet{Reis2014} is used here.

\begin{figure}
  \centering
  \includegraphics[width=\textwidth]{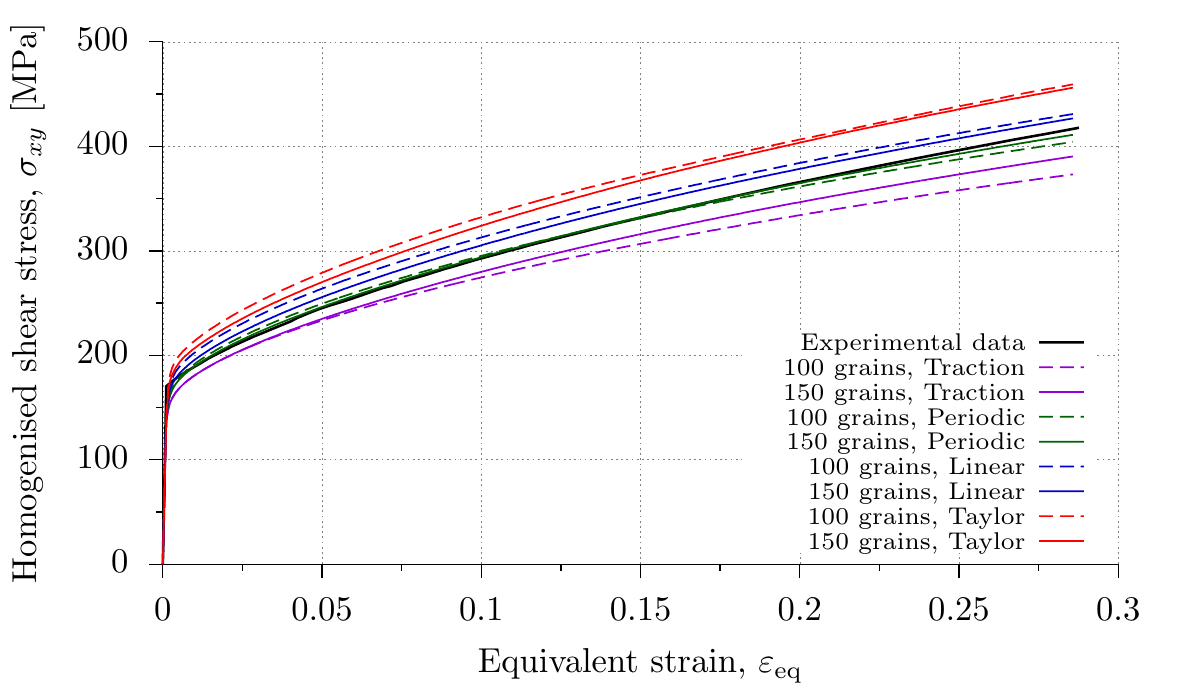}
  \caption{Homogenised shear stress-strain curves for stable austenite simple shear test. Results for both 100 and 150 grain RVEs are shown, for all applied homogenisation boundary conditions. Experimental data from \cite{Perdahcioglu2012}.}
  \label{fig:austenite-stress-strain}
\end{figure}

\begin{figure}
  \centering
  \begin{subfigure}{0.43\textwidth}
    \centering
    \includegraphics[width=\textwidth]{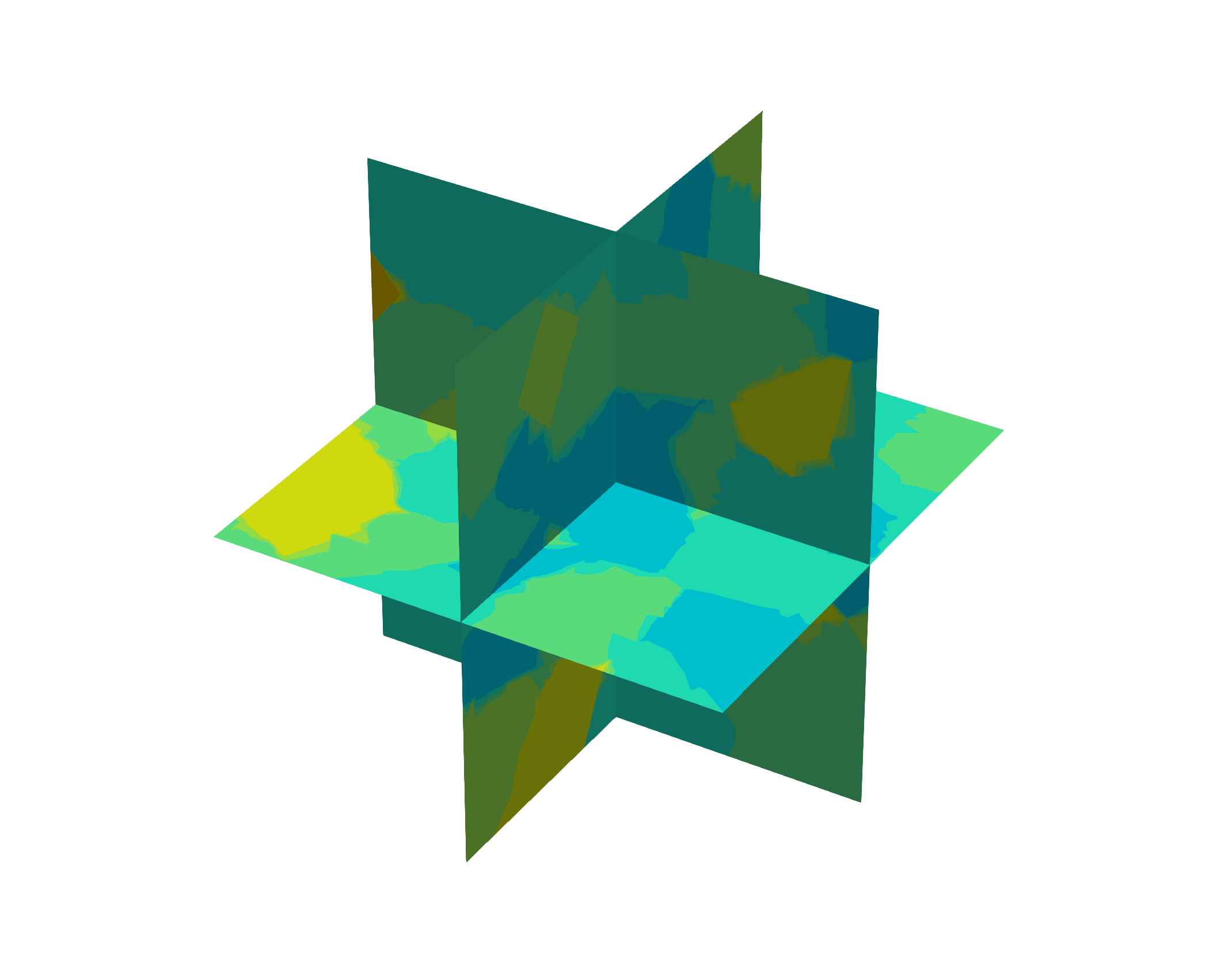}
    \includegraphics[width=\textwidth]{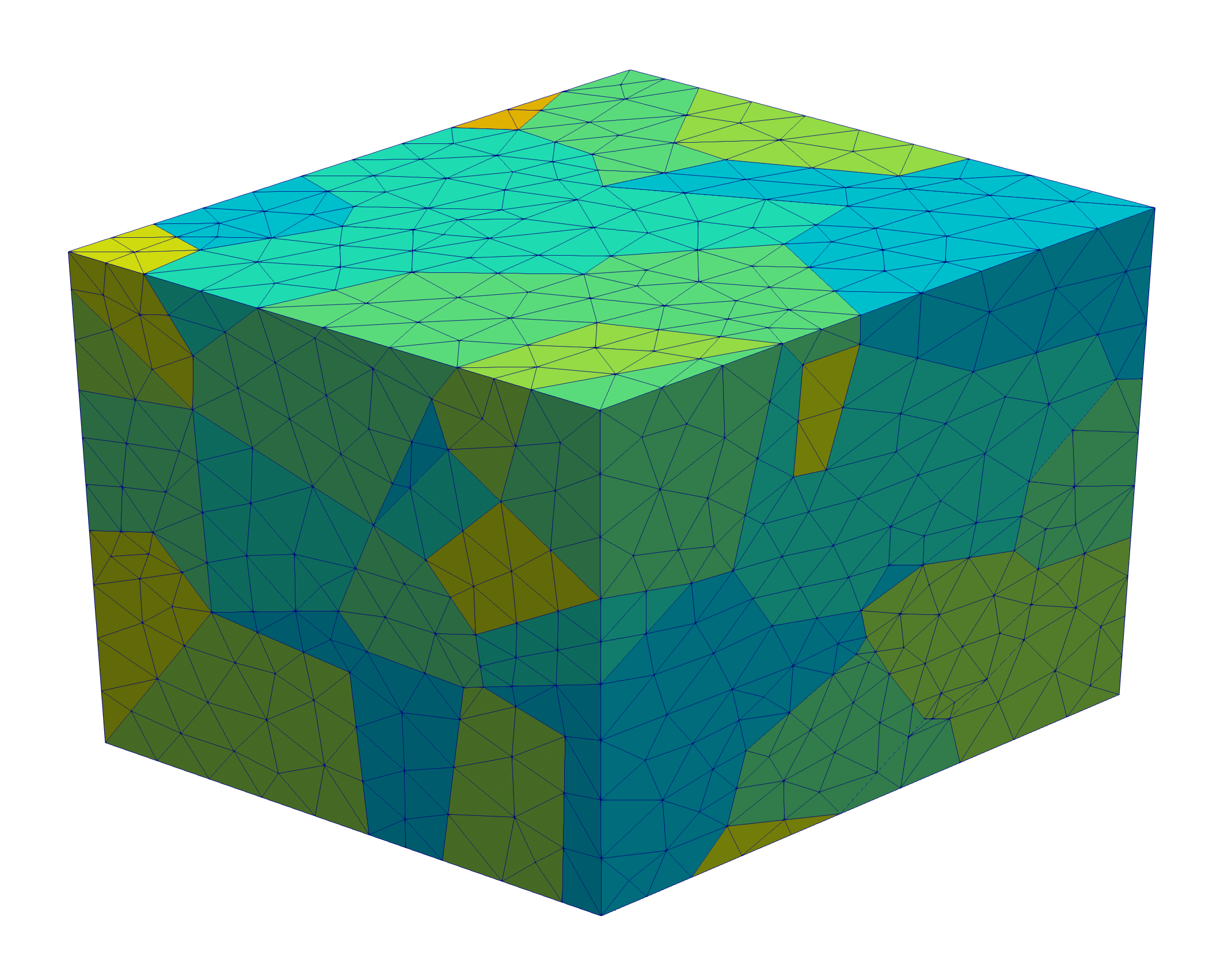}
    \caption{Taylor}
  \end{subfigure}
  \quad
  \begin{subfigure}{0.43\textwidth}
    \centering
    \includegraphics[width=\textwidth]{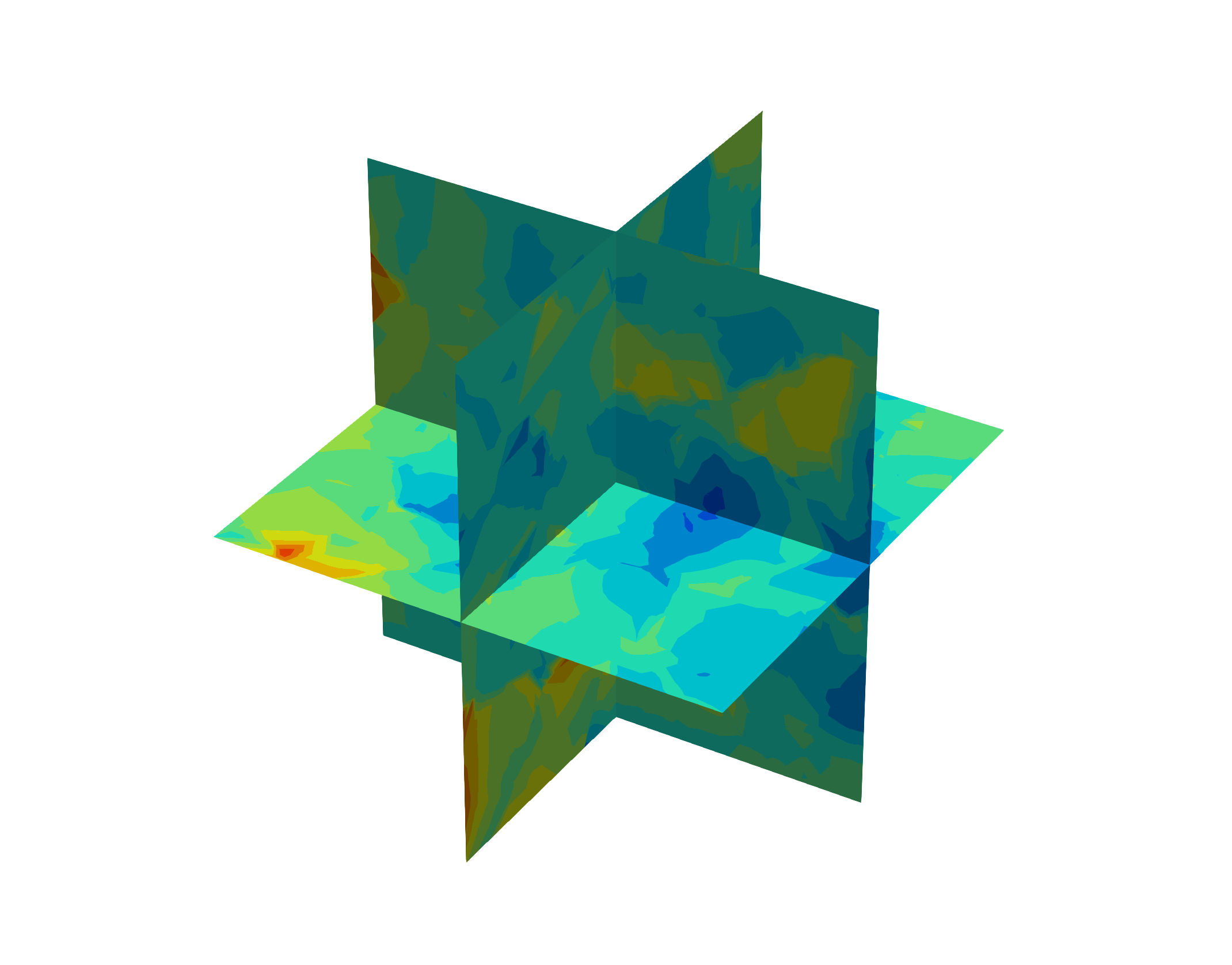}
    \includegraphics[width=\textwidth]{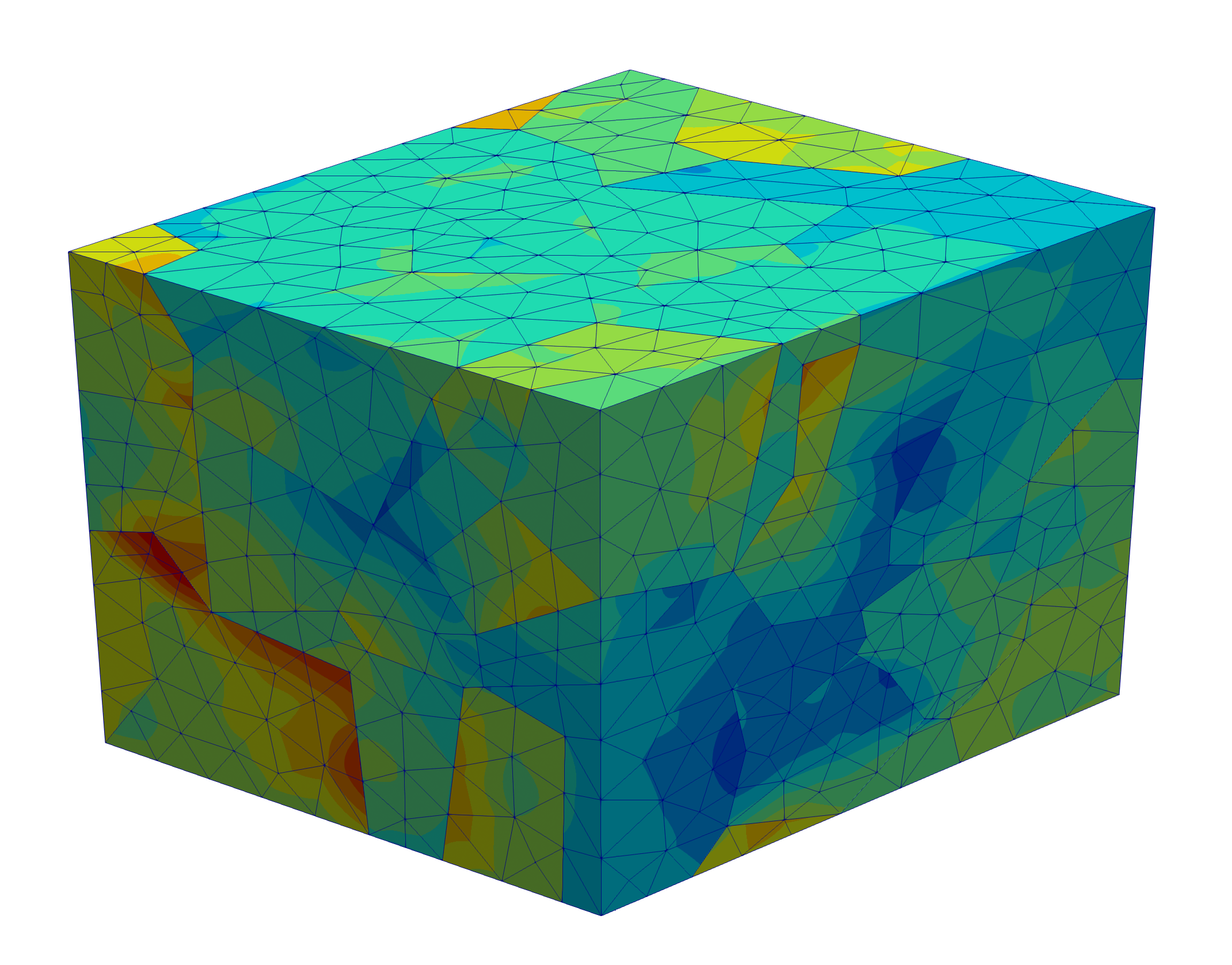}
    \caption{Linear}
  \end{subfigure}
  \par\bigskip
  \begin{subfigure}{0.43\textwidth}
    \centering
    \includegraphics[width=\textwidth]{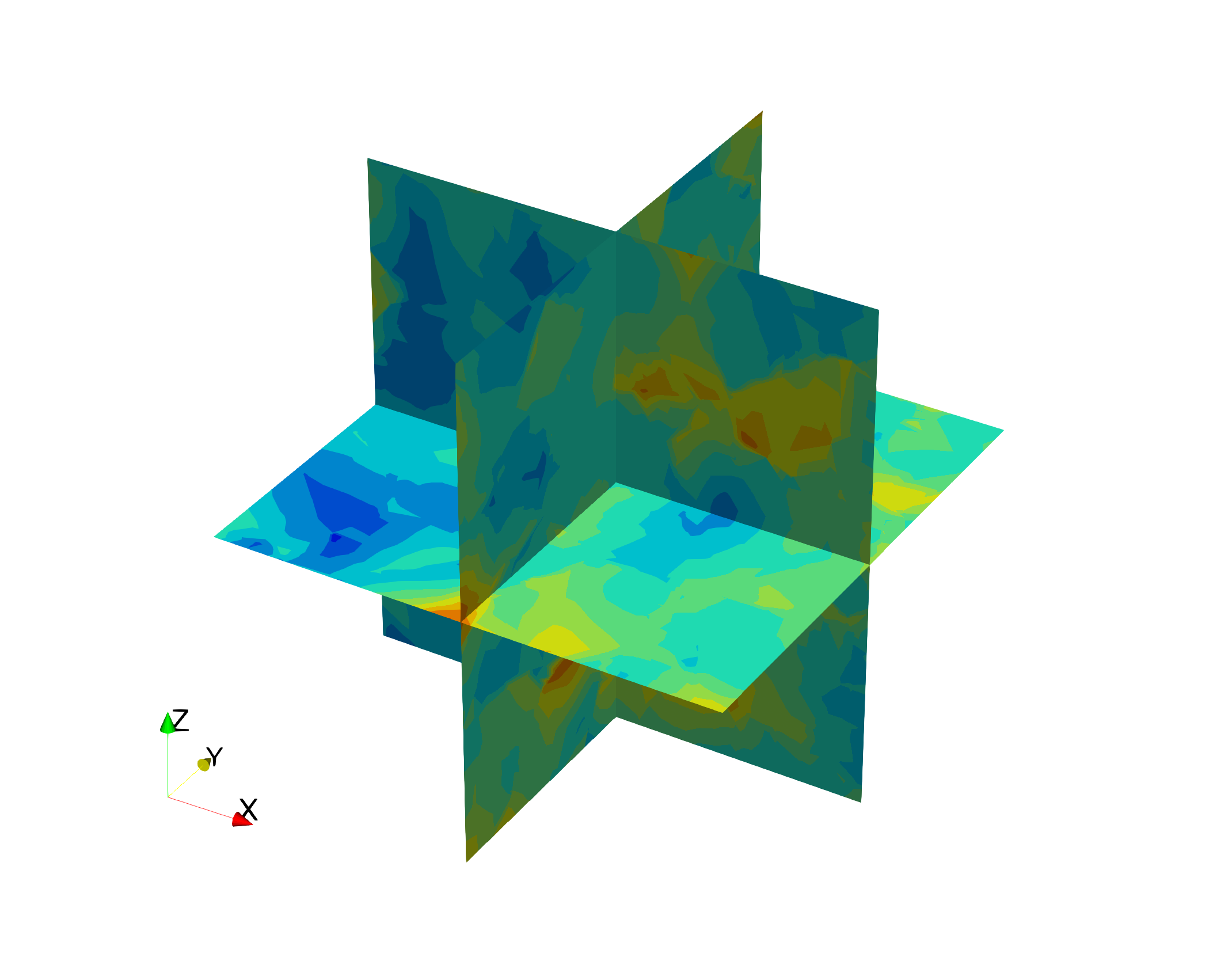}
    \includegraphics[width=\textwidth]{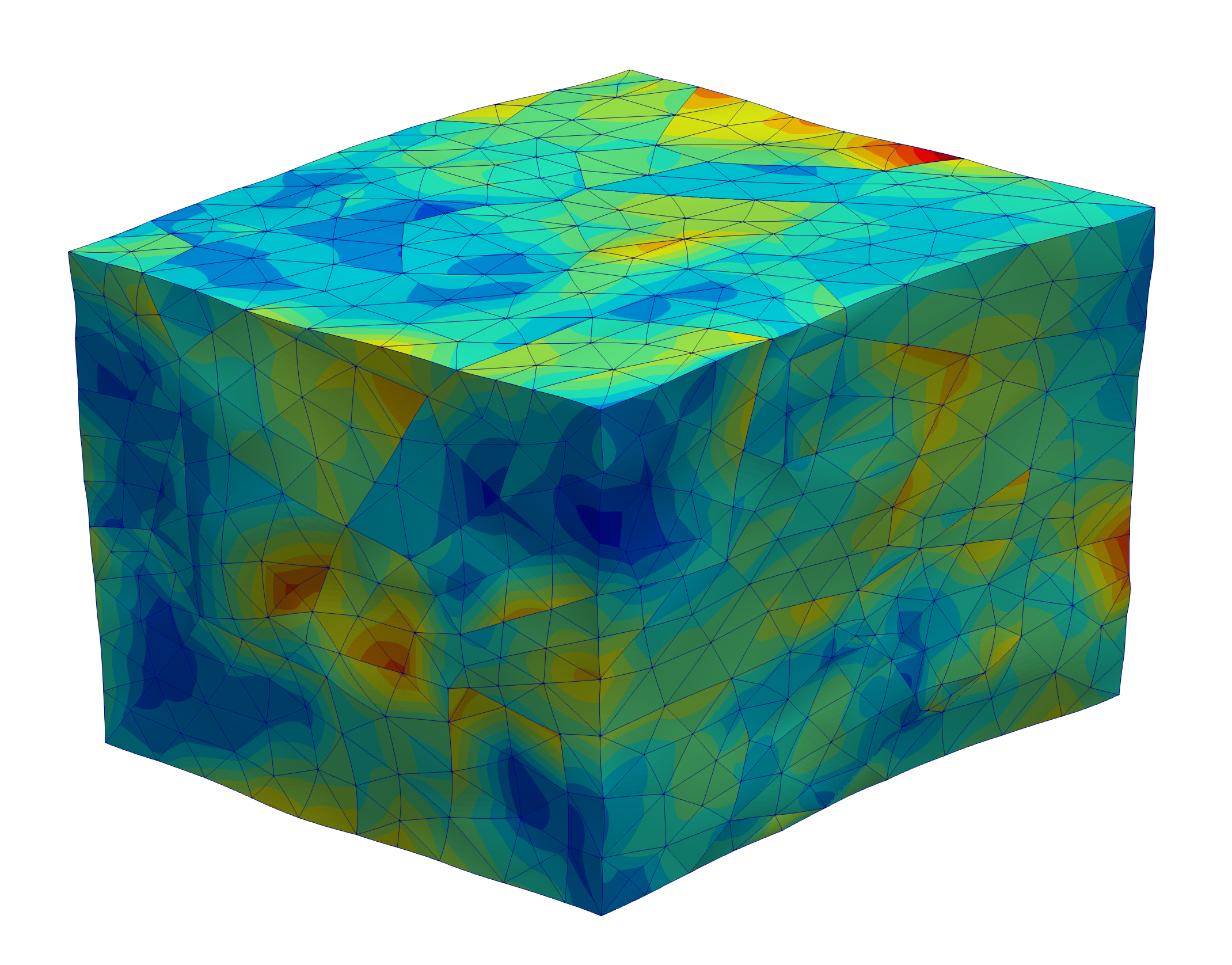}
    \caption{Periodic}
  \end{subfigure}
  \quad
  \begin{subfigure}{0.43\textwidth}
    \centering
    \includegraphics[width=\textwidth]{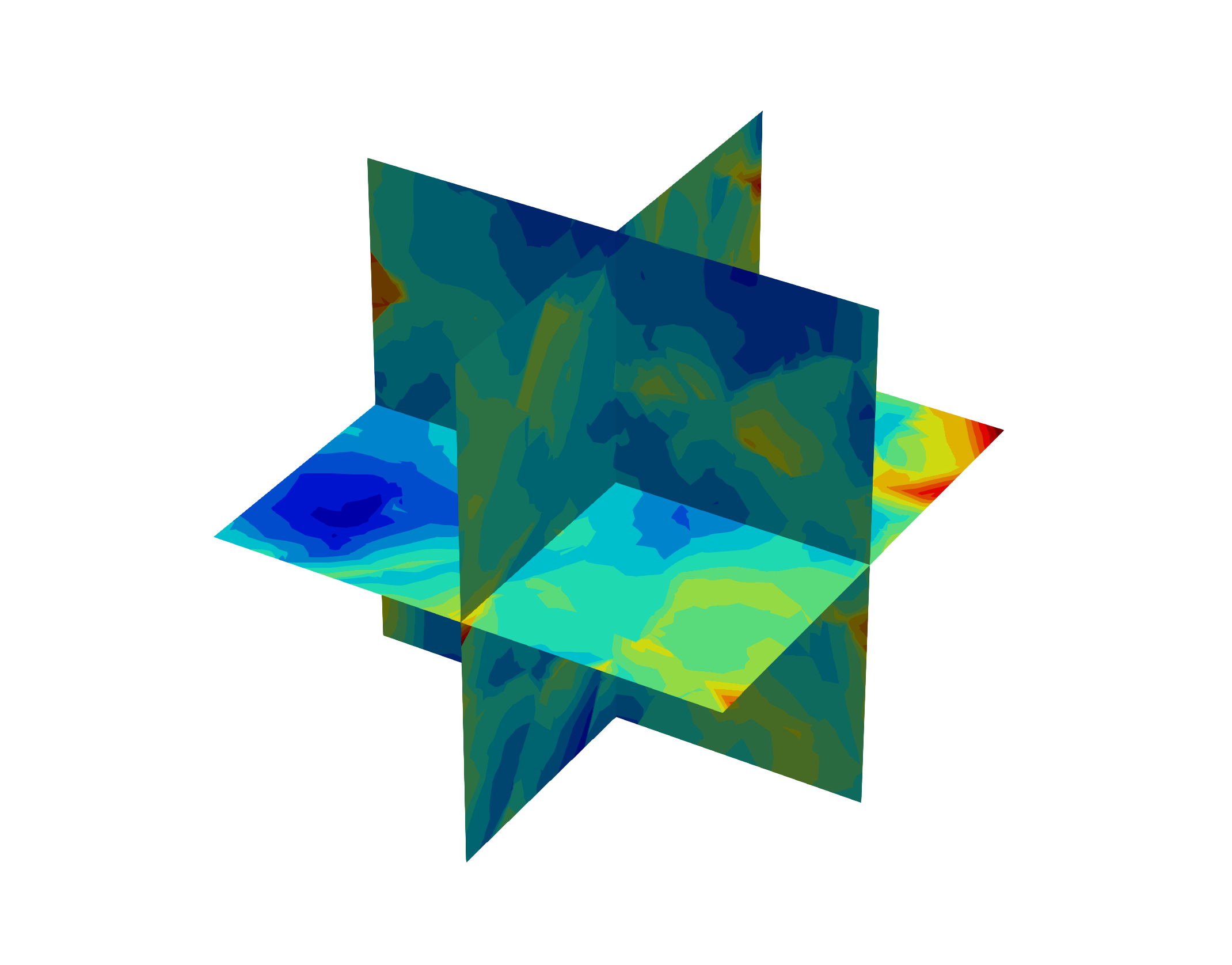}
    \includegraphics[width=\textwidth]{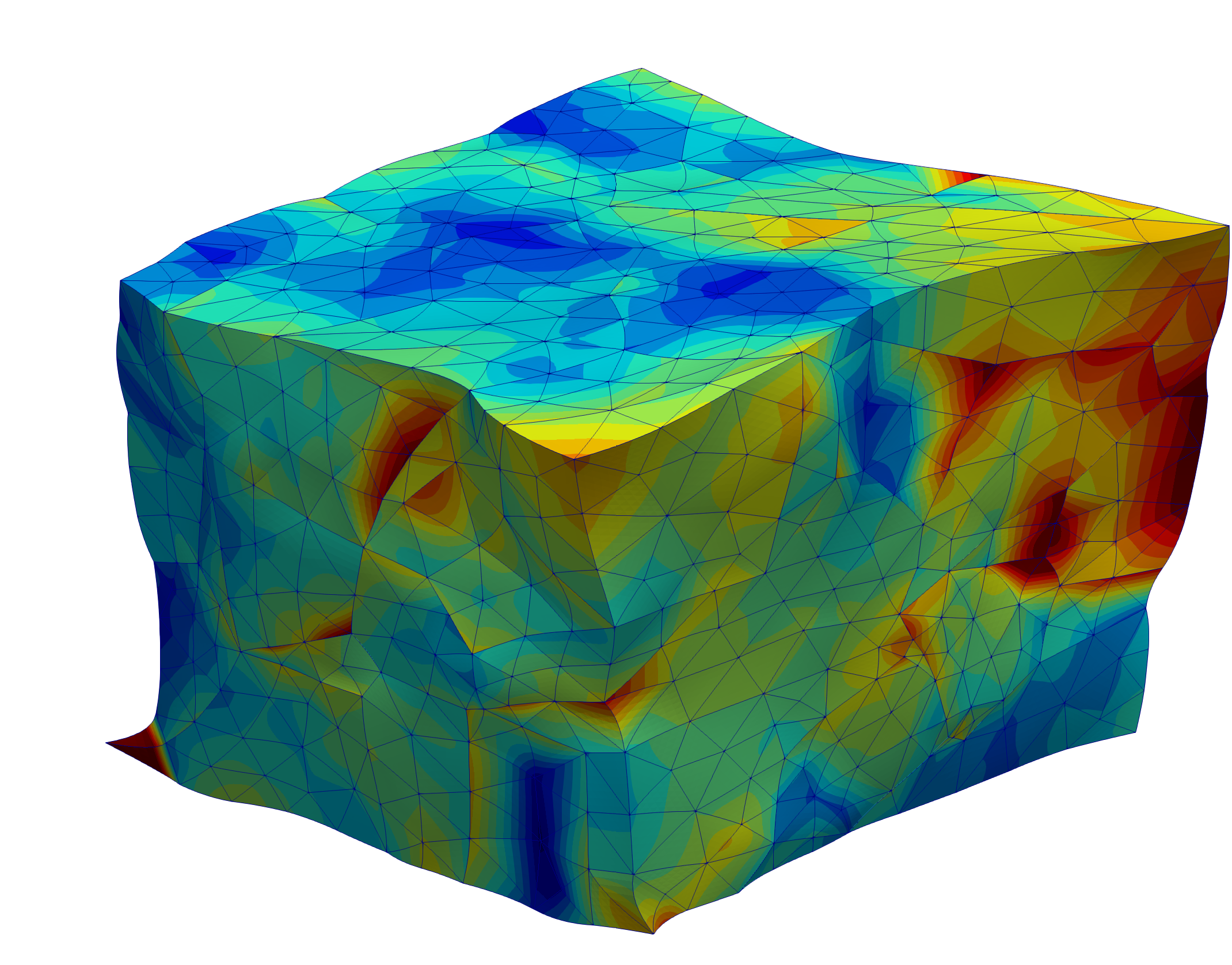}
    \caption{Uniform traction}
  \end{subfigure}
  \begin{subfigure}{\textwidth}
    \includegraphics[width=\textwidth]{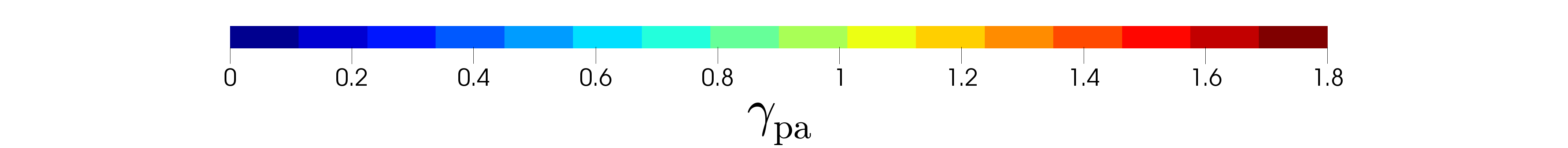}
  \end{subfigure}
  \caption{Accumulated plastic strain distribution at the end of the austenite $xy$-shear test for 100 grain RVE: deformed meshes and cuts showing intra-granular variable distribution.}
  \label{fig:100grains-deformed-austenite}
\end{figure}

\begin{figure}
  \centering
  \begin{subfigure}{0.43\textwidth}
    \centering
    \includegraphics[width=\textwidth]{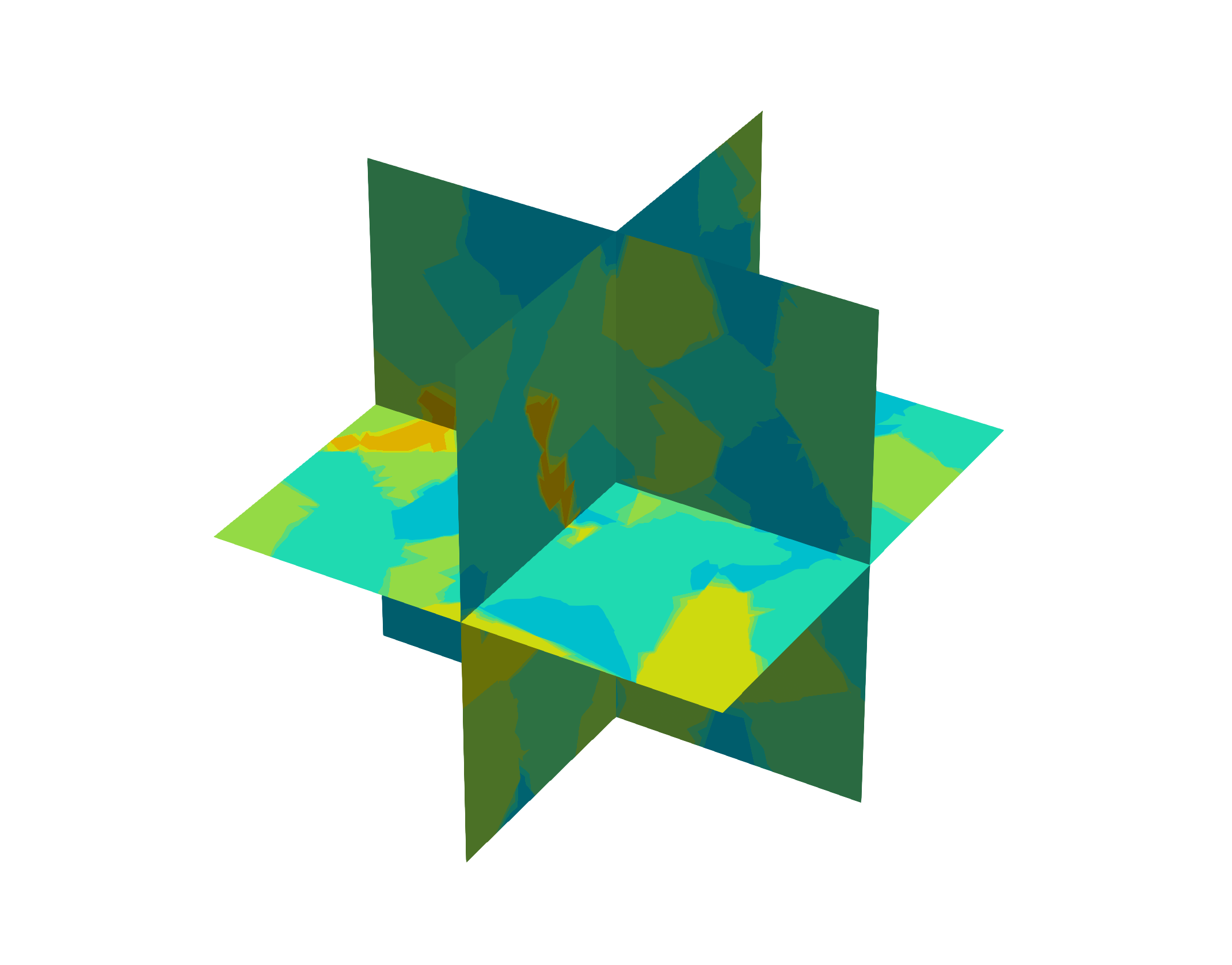}
    \includegraphics[width=\textwidth]{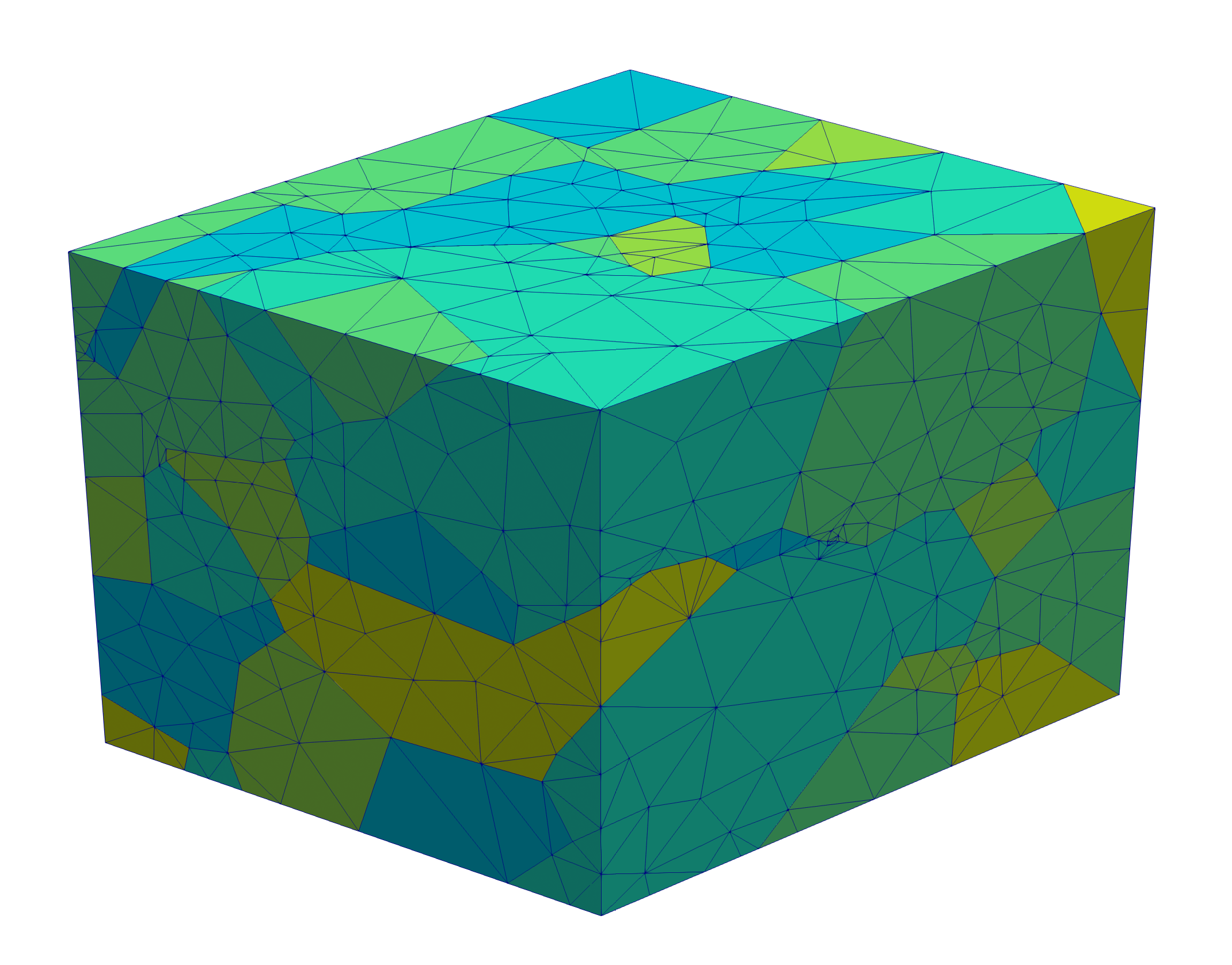}
    \caption{Taylor}
  \end{subfigure}
  \quad
  \begin{subfigure}{0.43\textwidth}
    \centering
    \includegraphics[width=\textwidth]{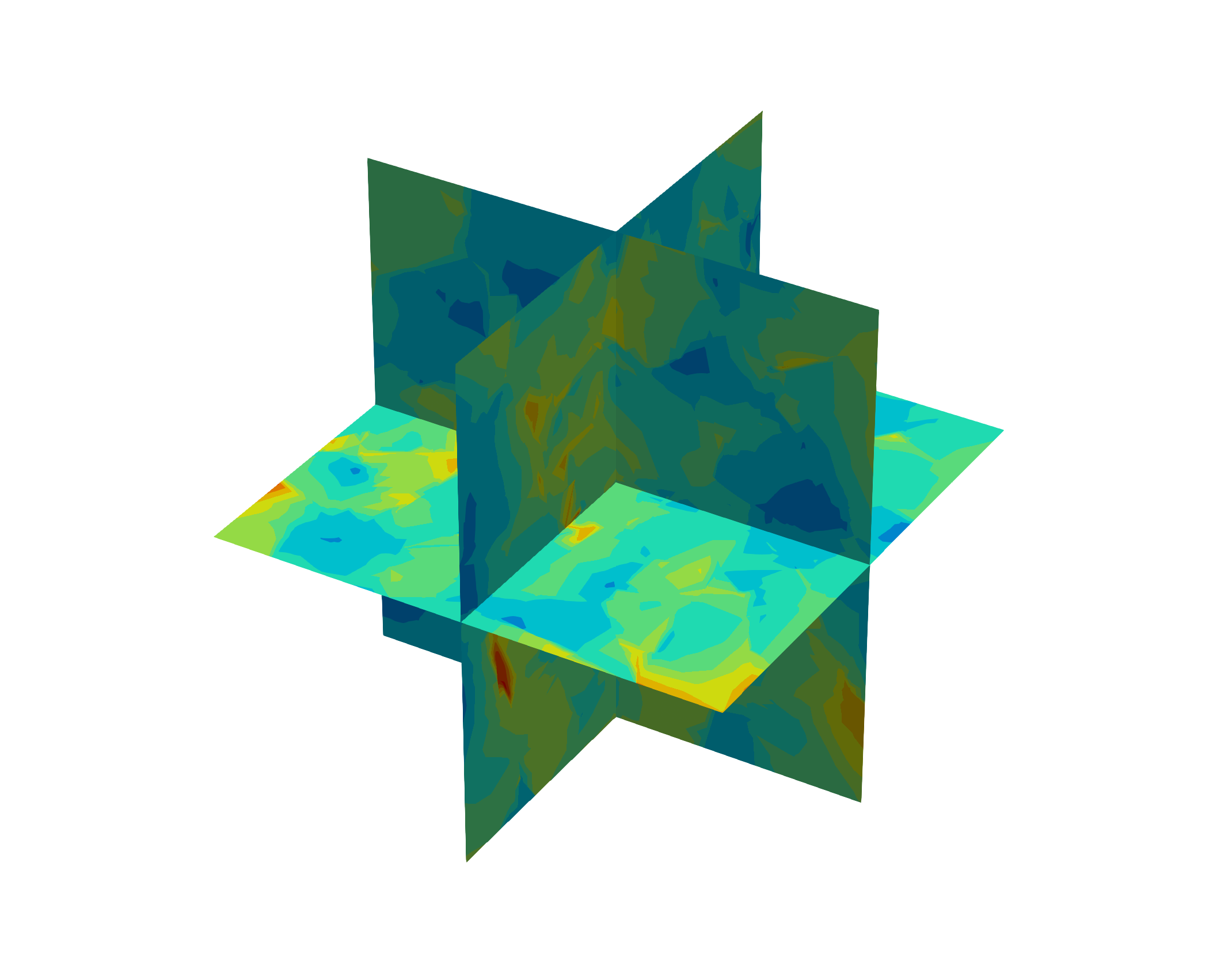}
    \includegraphics[width=\textwidth]{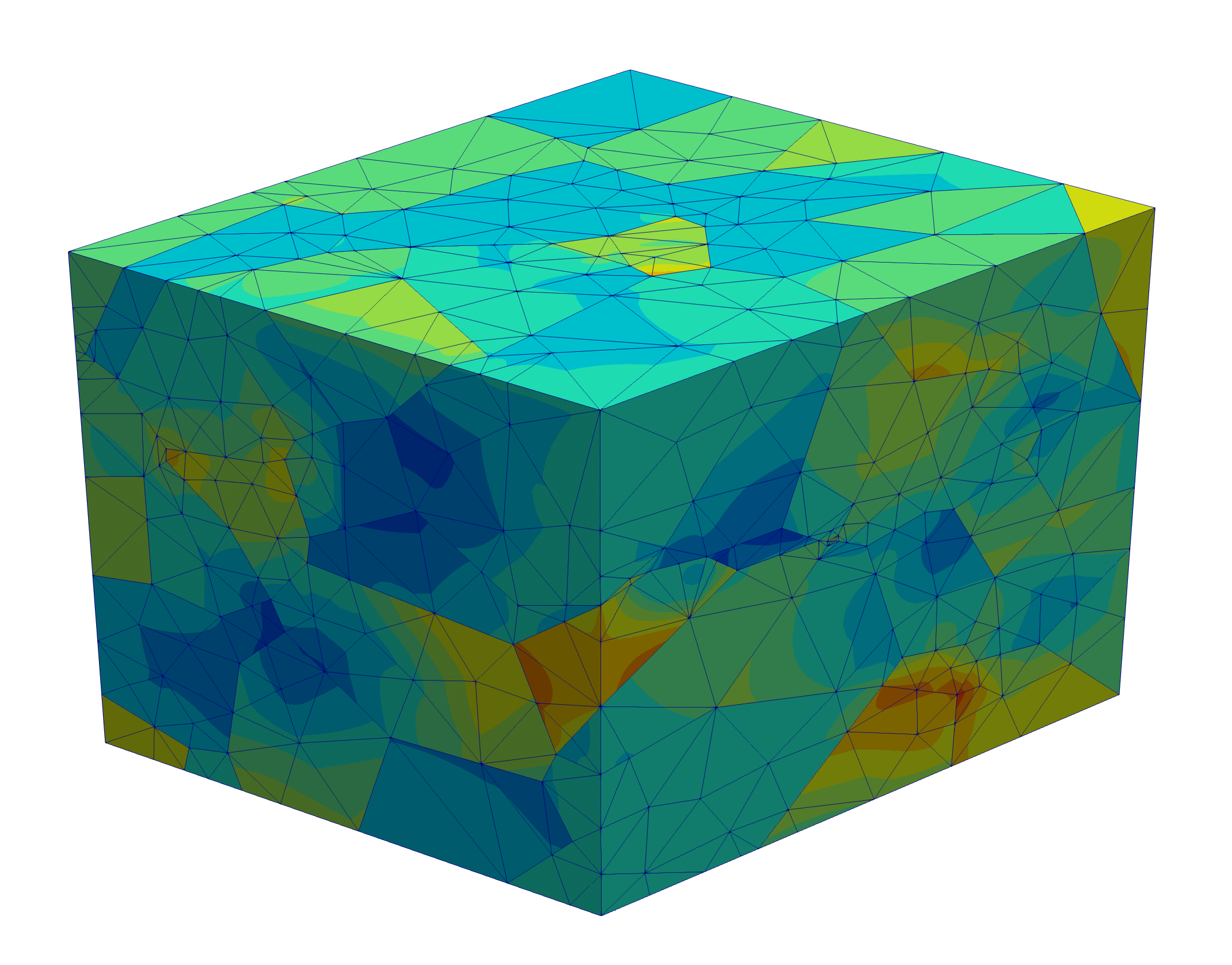}
    \caption{Linear}
  \end{subfigure}
  \par\bigskip
  \begin{subfigure}{0.43\textwidth}
    \centering
    \includegraphics[width=\textwidth]{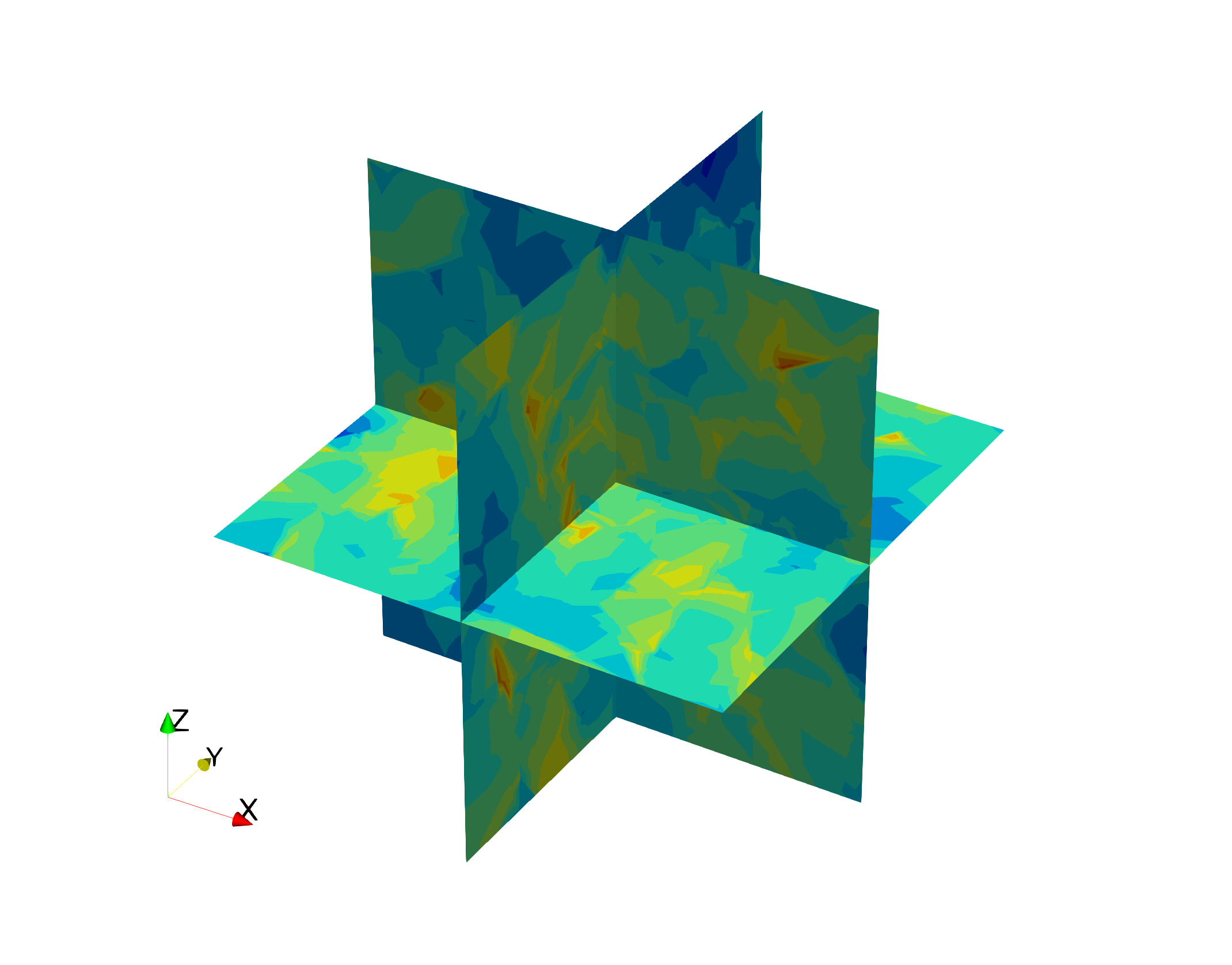}
    \includegraphics[width=\textwidth]{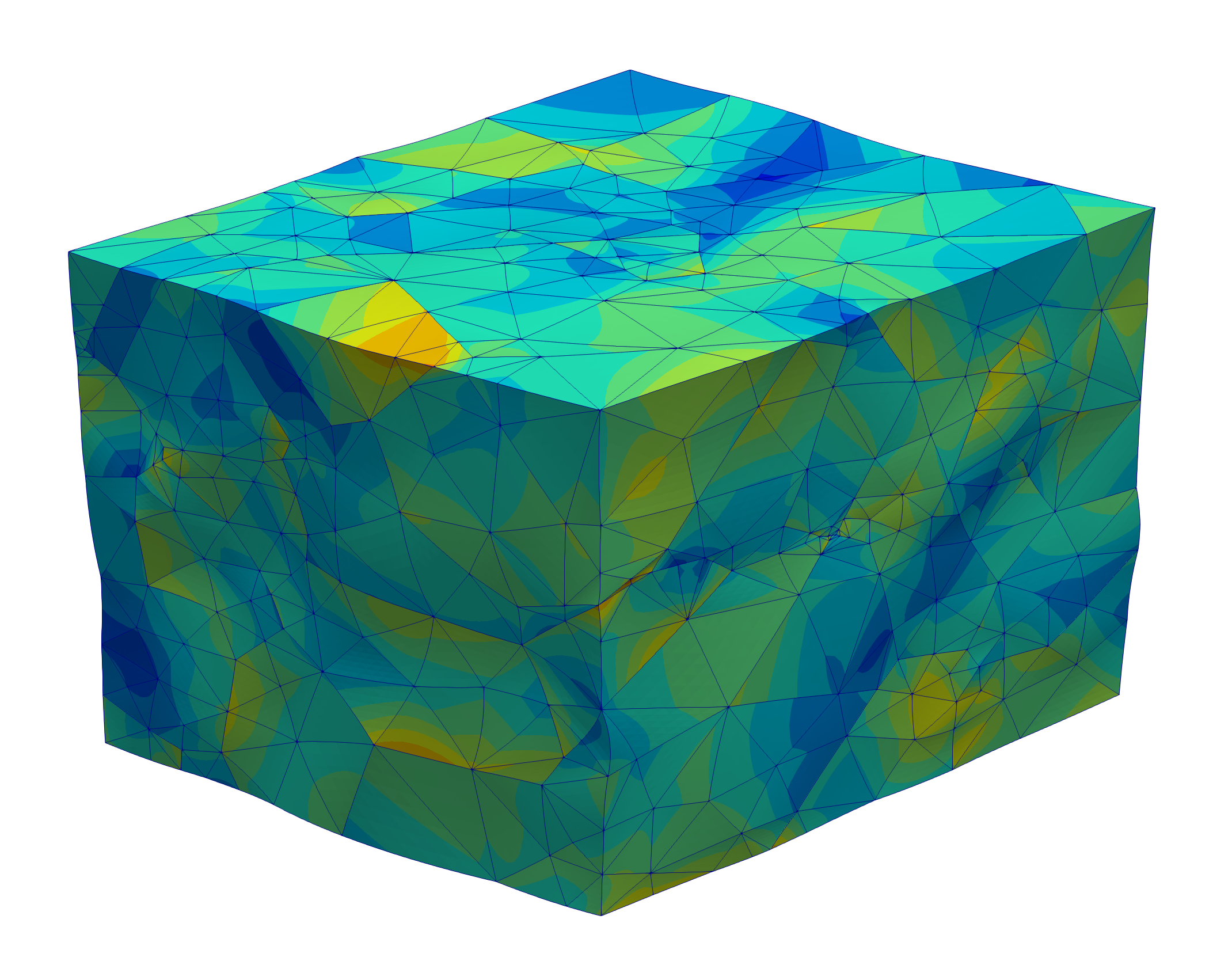}
    \caption{Periodic}
  \end{subfigure}
  \quad
  \begin{subfigure}{0.43\textwidth}
    \centering
    \includegraphics[width=\textwidth]{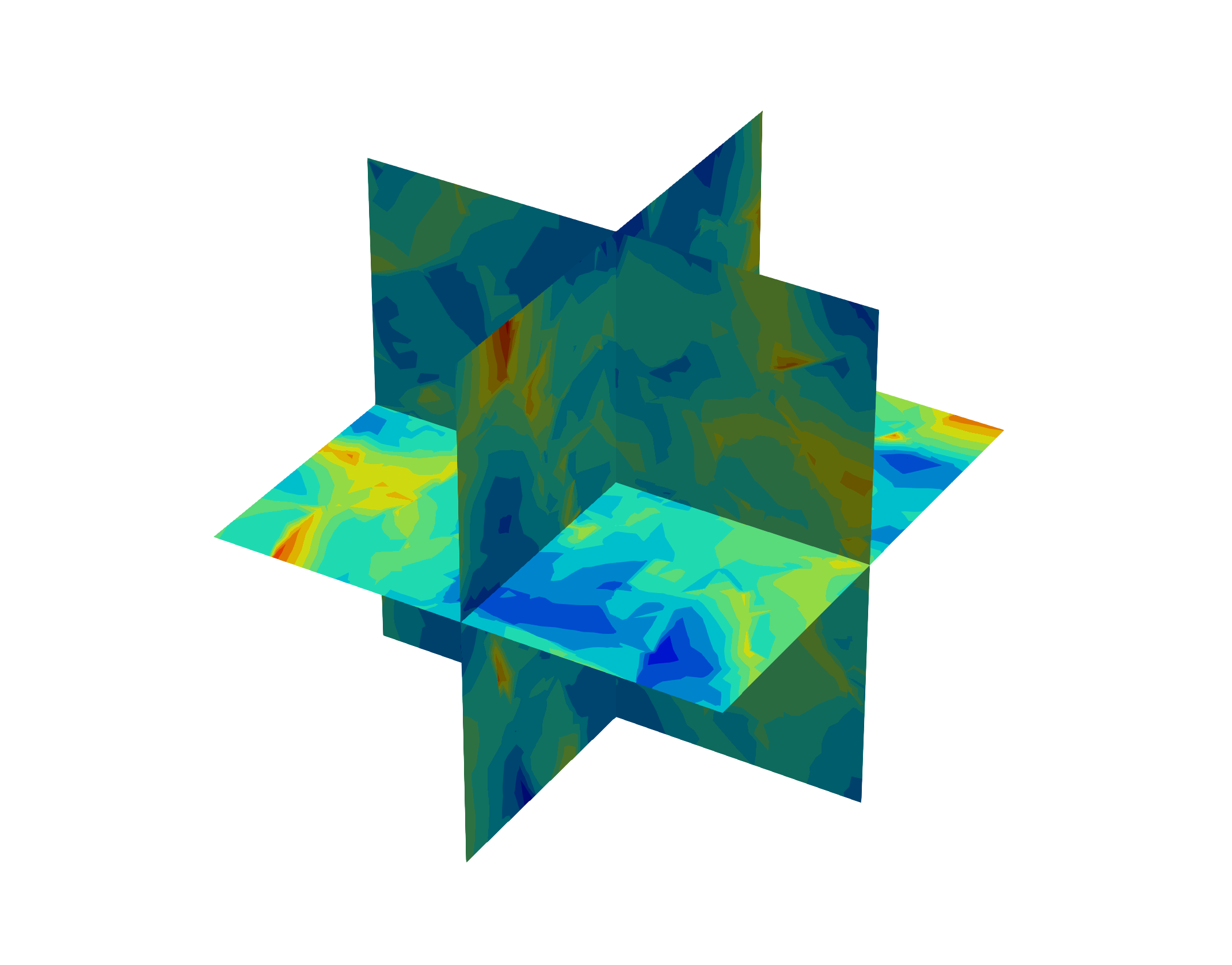}
    \includegraphics[width=\textwidth]{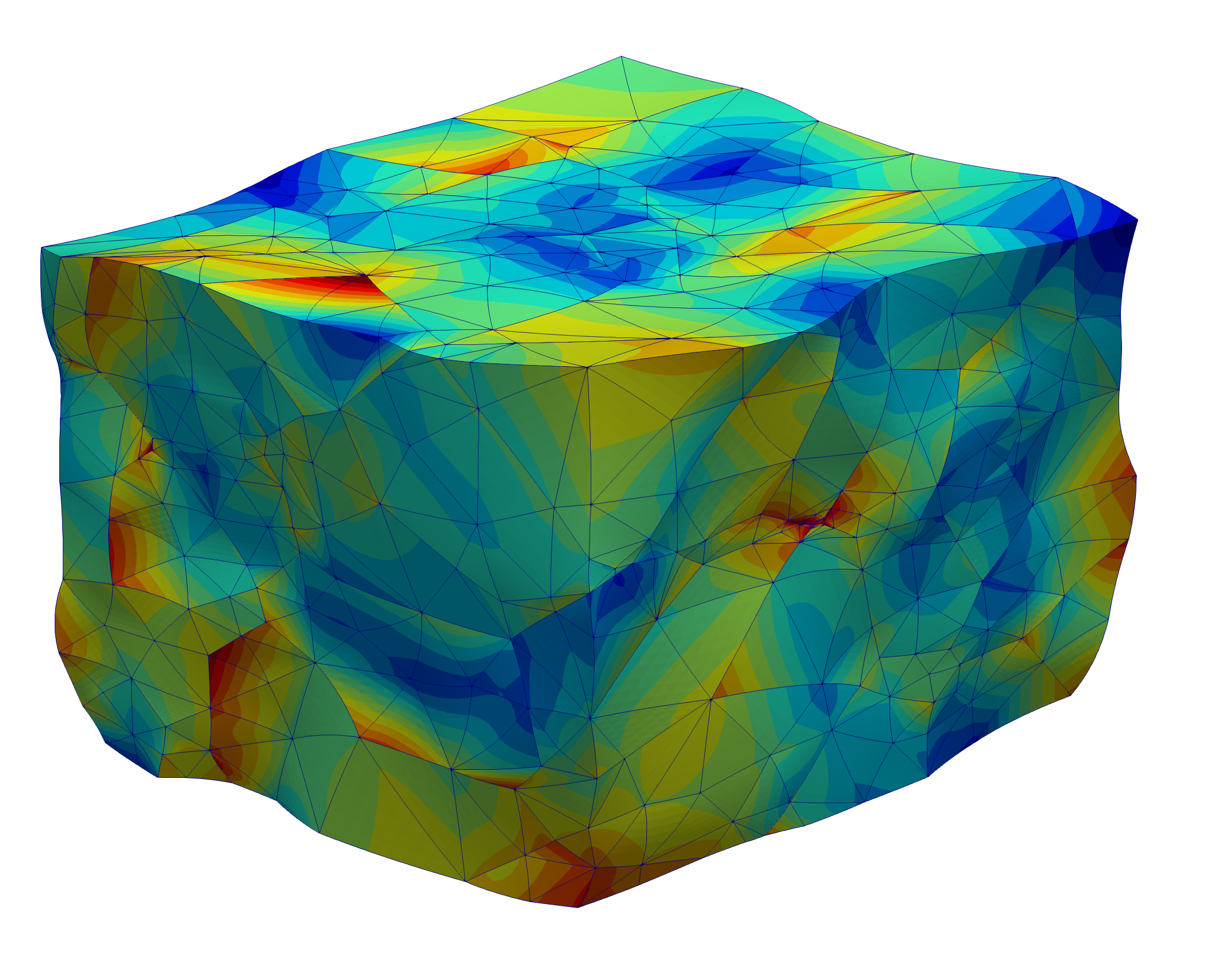}
    \caption{Uniform traction}
  \end{subfigure}
  \begin{subfigure}{\textwidth}
    \includegraphics[width=\textwidth]{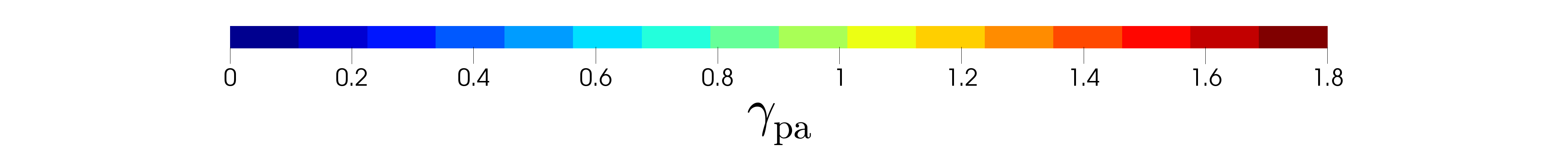}
  \end{subfigure}
  \caption{Accumulated plastic strain distribution at the end of the austenite $xy$-shear test for 150 grain RVE: deformed meshes and cuts showing intra-granular variable distribution.}
  \label{fig:150grains-deformed-austenite}
\end{figure}

The resulting homogenised stress-strain curves are shown in \cref{fig:austenite-stress-strain} for both meshes and all boundary conditions. From these results, it can be seen that the homogenised material behaviour is practically the same for both RVEs, confirming that 150 is a large enough number of grains to capture the isotropic material behaviour. Thus, in subsequent analyses, only the 150 grain RVE will be used, as the computational cost associated to both meshes is approximately the same (as they have similar numbers of degrees of freedom) and no appreciable difference can be expected from adding more crystallites. \Cref{fig:100grains-deformed-austenite,fig:150grains-deformed-austenite} show the accumulated plastic slip distribution at the final deformation stage for both RVEs and all boundary conditions. It can be readily noticed that although both RVEs behave similarly from a homogenised perspective, their internal variable and displacement fluctuation distributions are completely different.

\FloatBarrier
\subsection{Homogenised stress-strain curves}

\begin{figure} [h]
  \centering
  \includegraphics[width=\textwidth]{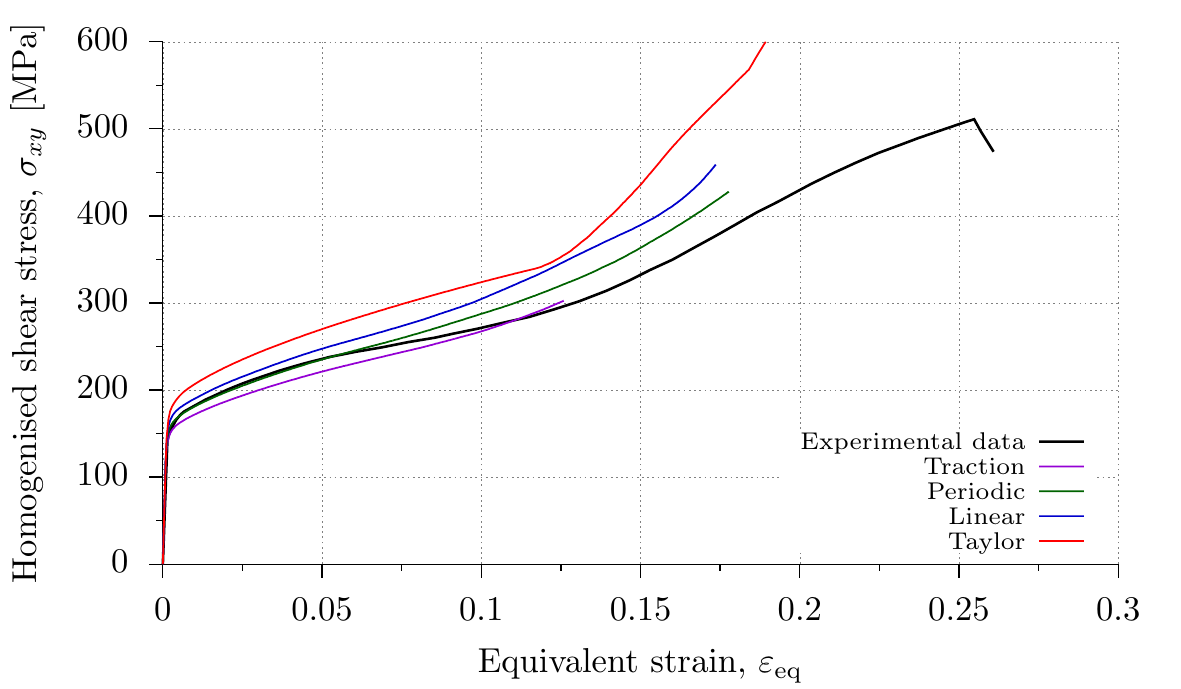}
  \caption{Homogenised shear stress-strain curves for a simple shear test, showing both the current model's predictions and experimental data from \cite{Perdahcioglu2012}.} \label{fig:transformation-stress-strain}
\end{figure}

Here, the full martensitic transformation material model is tested in the same simple shear example. With the calibrated hardening curve for the austenite at hand, the RVE is loaded once again in a shear deformation of 50\% with the martensitic transformation enabled. The resulting homogenised stress-strain curves are shown in \cref{fig:transformation-stress-strain}, along with experimental data from \cite{Perdahcioglu2012}. It can be seen that the model shows good agreement with the experimental results, capturing the observed hardening effect due to the transformation. However, an overly stiff response is observed once the transformation begins; this can be interpreted as a result of the simple coupling approach between transformation and austenite slip plasticity since, at a given material point, the latter stops once the former begins. Additionally, as the deformations become large, numerical convergence difficulties ensue due to the stiffness of the regularised model equations.

\FloatBarrier
\subsection{Transformation surface}

To produce a transformation surface visualisation, the RVE is loaded with a set of macroscopic deformation gradients $\tenII{F}$ representing different loading paths. For each deformation path, the homogenised Cauchy stress at which transformation starts at any material point is recorded, providing a point of the transformation surface. Rather than working with a hyper-surface in six-dimensional stress space, for which a visualisation would be impossible, it is instead preferable to use the three-dimensional principal Cauchy stress space -- also known as \emph{Haigh-Westergaard} space -- with coordinates $\sigma_1$, $\sigma_2$ and $\sigma_3$. Even in this case, instead of providing a full three-dimensional depiction of the surface, its projections in the deviatoric stress plane\footnote{The deviatoric stress plane is the set of principal Cauchy stresses that satisfy $\sigma_1 + \sigma_2 + \sigma_3 = 0$.}, at different levels of hydrostatic pressure $p = (\sigma_1 + \sigma_2 + \sigma_3)/3$, provide a more convenient visualisation.

\FloatBarrier
\subsubsection{Deviatoric plane cross section}

\begin{figure} [h]
  \centering
  \includegraphics[width=\textwidth]{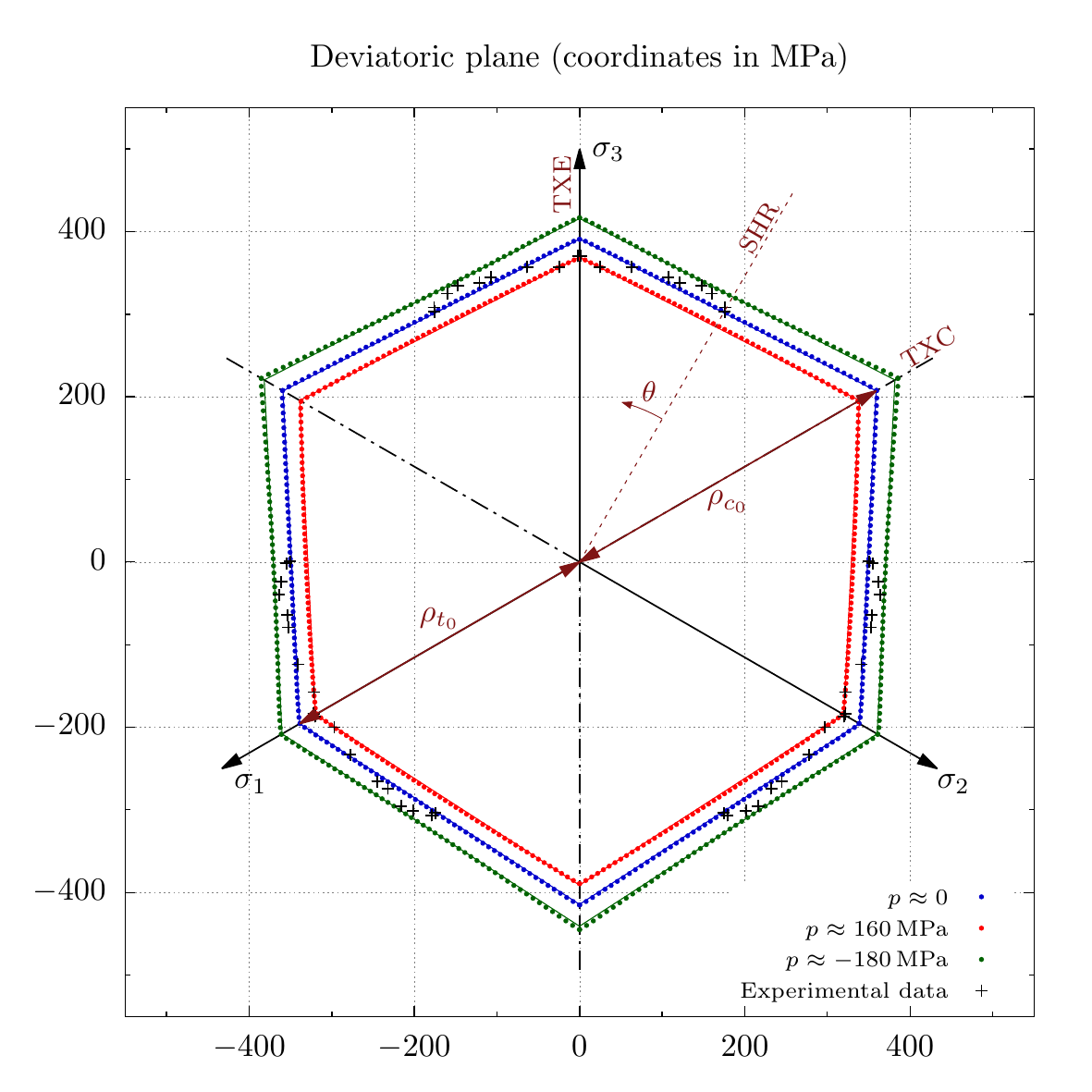}
  \caption{Transformation surface in deviatoric plane at different pressure levels, highlighting the pressure-dependent nature of the transformation criterion. Positive (tensile) hydrostatic pressures favour the transformation onset, while negative (compressive) ones have the opposite effect. Continuous curves represent Mohr--Coulomb surface corresponding to this material's transformation parameters at the different hydrostatic pressures. Lode angle $\theta$ and stress states corresponding to simple shear (SHR), triaxial extension (TXE) and triaxial compression (TXC) are also labelled. Experimental data from \cite{Geijselaers2009}.} \label{fig:transformation-surface}
\end{figure}

To cover all directions in the deviatoric stress plane, a loading programme ranging from triaxial extension (two equal principal stresses with the third one being higher) to simple shear (one principal stress equal to zero, and the other two equal but opposite in sign) and finishing at triaxial compression (analogous to triaxial extension but with the third eigenvalue lower than the other two) has to be devised. These stress states are labelled, respectively, TXE, SHR and TXC in \cref{fig:transformation-surface} and correspond to Lode angles $\theta$ of +\SI{30}{\degree}, \SI{0}{\degree} and -\SI{30}{\degree}, with
\begin{equation}
  \theta = \frac{1}{3} \arcsin \left[ \frac{J_3}{2} \left( \frac{3}{J_2} \right)^{3/2}  \right],
\end{equation}
where $J_2 \equiv  \frac{1}{2}\tenII{s}:\tenII{s}$, $J_3 \equiv \det \left[ \tenII{s} \right]$ and $\tenII{s} \equiv \dev\left[ \tenII{\sigma} \right]$.

As the transformation criterion is expected to be pressure-dependent and the computational code in which these results are obtained is strain-driven -- that is, in the case of a homogenisation analysis, the deformation gradient is the main input variable and the homogenised stress tensor the main output -- it is not possible to directly impose a given stress state on the RVE. In particular, given the pressure-independent nature of the austenite slip plasticity constitutive model, even if care is taken to impose a set of deformation gradients with equal volumetric component (i.e. same determinant), the resulting homogenised hydrostatic pressure is not the same for each loading path. Thus, it is not possible to directly obtain a cross section of the transformation surface at a given hydrostatic pressure level using the full transformation and plasticity model.

A simple way to circumvent that limitation involves the suppression of the austenite plastic deformation, so that all grains in the polycrystal behave elastically until the transformation starts. In this case, the homogenised hydrostatic pressure is found to be approximately equal for all imposed macroscopic deformation gradients with a given determinant. One drawback of this approach is due to the isotropic hyperelastic potential adopted in this model: all grains deform identically until the transformation starts, as the crystallographical orientation has no effect until that point. This means that the different homogenisation boundary conditions yield the same result; with that in mind, results in this section will only be presented for the periodic boundary condition.

To impose the load programme on the RVE, a one-parameter family of deformation gradients $\tenII{F}(\alpha)$ is defined as a function load parameter $\alpha \in [-1, 1]$. Ideally, the resulting Lode angle should vary linearly with $\alpha$, so that a minimal number of analysis points will yield a meaningful transformation surface cross section. One such family can be defined by:
\begin{equation}
  \tenII{F}(\alpha) = \begin{bmatrix}
                        1 + \alpha\eta &  (1-\alpha^2)\eta   & 0                                                   \\
                        0              &   1 + \alpha\eta    & 0                                                   \\
                        0              &     0               & \displaystyle\frac{1}{\left(1+\alpha\eta \right)^2}
                      \end{bmatrix},
\end{equation}
which is isochoric for all values of $\alpha$, to evaluate the transformation surface at $p \approx 0$. For a given value of $\alpha$, the load factor $\eta$ is defined so that the equivalent strain $\varepsilon_\text{eq}$ is equal to a pre-defined value $\overbar{\varepsilon}$. In general, $\eta$ varies as function of $\alpha$, so a simple iterative procedure based on the bisection method is used to find it for each prescribed deformation gradient.

\begin{figure} [h]
  \centering
  \includegraphics[width=\textwidth]{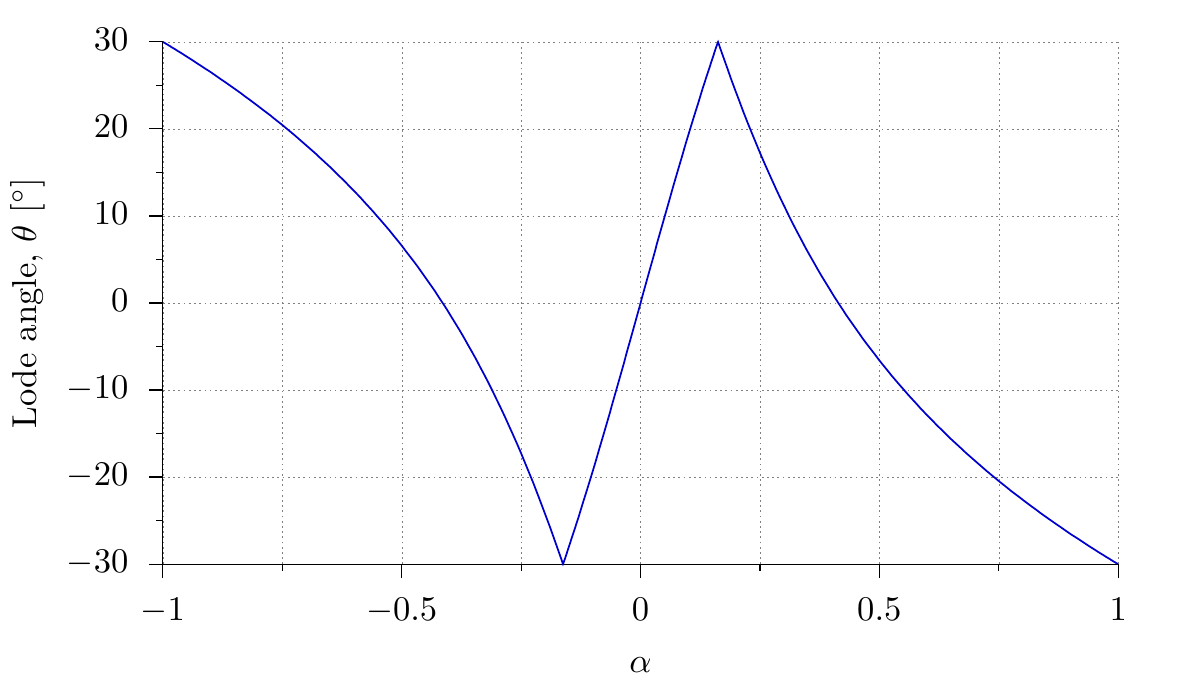}
  \caption{Lode angle as a function of deformation gradient parameter $\alpha$.} \label{fig:lode-angle}
\end{figure}

To predict the Lode angles resulting from the prescription of this deformation gradient family for $\alpha \in [-1, 1]$, its value is calculated analytically assuming a linear elastic material deforming under small strains, as a first approximation. In this case, the principal stresses are directly proportional to the principal strains, so the Lode angle can be easily determined. The results are displayed in \cref{fig:lode-angle}, clearly showing that the full range of Lode angles is covered. However, it can be seen by restricting the value of $\alpha$ to the interval $[-0.163, 0.163]$, the full range of Lode angles $\theta \in [-\SI{30}{\degree}, \SI{30}{\degree}]$ is also covered with no overlap and in almost linear fashion. The limits of this interval ($\alpha \approx \pm 0.163$) are determined by finding the values of $\alpha$ for which two of the principal stresses are equal, with the third one being non-zero.

With the loading programme defined, the RVE is deformed along each prescribed $\tenII{F}$ until the transformation starts at any material point in the mesh; at that moment, the principal Cauchy stresses ${\sigma_1, \,\sigma_2, \,\sigma_3}$ are recorded. Assuming that these are ordered such that $\sigma_1 \geq \sigma_2 \geq \sigma_3$, their values are used to compute the deviatoric plane coordinates of a point in the transformation surface\footnote{Since it is the isotropic limit of the material behaviour that is of interest, the orientation used to describe the macroscopic stresses has no significance.}. Varying $\alpha$ in the predefined range, a sextant of the transformation surface's deviatoric plane projection is covered. The other five are then obtained by appropriate permutations of $\sigma_1$, $\sigma_2$ and $\sigma_3$, resulting in \cref{fig:transformation-surface}. Data reported by \cite{Geijselaers2009} are also shown, from which it is clear that the martensitic transformation model shows very good agreement with the experiments. The deviatoric plane projection of the equivalent Mohr--Coulomb locus is also shown at all hydrostatic pressures analysed, showing once again excellent agreement with the numerical results.

To study the pressure-sensitivity of the transformation criterion, the same analyses are run two more times using non-isochoric deformation gradients. These are obtained from the same set $\tenII{F}(\alpha)$ with an added volumetric component, which is expansive in the first case ($\det\left[ \tenII{F} \right] > 1$) and compressive in the second ($\det\left[ \tenII{F} \right] < 1$). This yields the other two sets of points shown in \cref{fig:transformation-surface}: one with hydrostatic pressure $p \approx \SI{160}{\mega\pascal}$ and another with $p \approx -\SI{180}{\mega\pascal}$. As would be expected given the volumetric expansion component of the transformation in this material, positive (tensile) hydrostatic pressures favour the transformation, whereas negative (compressive) hydrostatic pressures inhibit its onset.

\subsubsection{Meridional profile}

\begin{figure} [h]
  \centering
  \includegraphics[width=\textwidth]{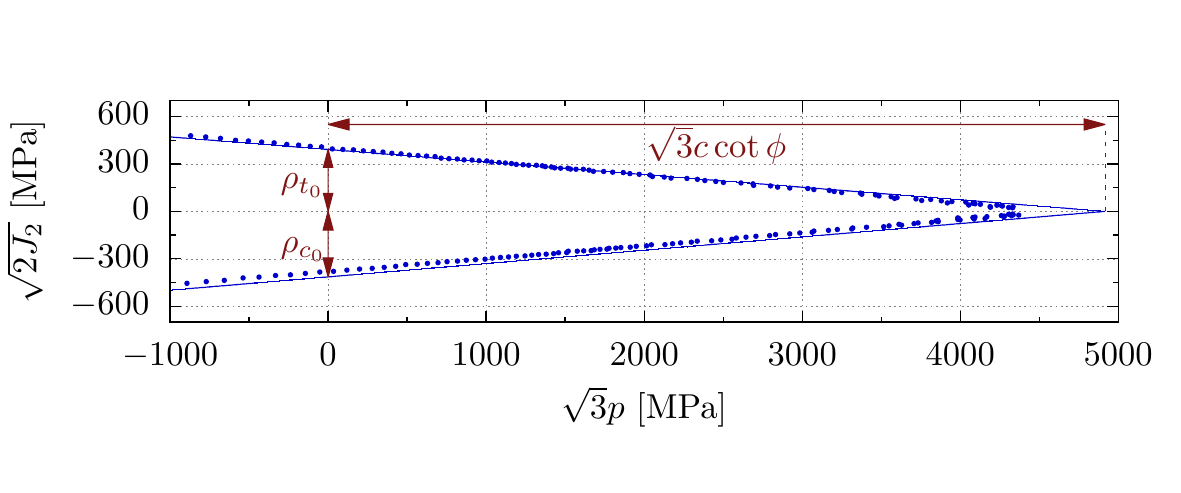}
  \caption{Meridional plane of transformation surface at $\theta = \SI{30}{\degree}$, showing its tensile-compressive asymmetry. Continuous curves represent Mohr--Coulomb surface corresponding to this material's transformation parameters.} \label{fig:meridional-plane}
\end{figure}

Another way of highlighting the pressure dependency of the surface is through its meridional profile: by holding the Lode angle constant and varying the hydrostatic pressure, a cross section of the surface along the hydrostatic axis (i.e. the axis along direction $(1, 1, 1)$ in principal stress space) is obtained. The results of one such series of analyses, for Lode angle $\theta = \SI{30}{\degree}$, are shown in \cref{fig:meridional-plane}, where Haigh--Westergaard coordinates are used to maintain isomorphism with principal stress space (i.e. lengths and angles are preserved between both representations).

Using \cref{eq:MC-params,eq:nanoflex-params}, the Mohr--Coulomb-like surface equivalent to this material is predicted to have the following parameters:
\begin{equation}
  \phi = \tan^{-1} \left(\frac{\delta}{\xi}\right) = \SI{4.980}{\degree}, \qquad
  c  = \frac{\DGm}{\xi} = \SI{247.4}{\mega\pascal}.
\end{equation}
The validity of this equivalence can be further checked using known properties of the Mohr--Coulomb yield surface, such as the fact the surface apex is located at a hydrostatic pressure value of:
\begin{equation}
  p = c \cot \phi = \frac{\DGm}{\delta} = \SI{2839}{\mega\pascal},
\end{equation}
which can readily be seen to agree with the results of \cref{fig:meridional-plane}. While this extremely high pressure is not likely to be relevant in practice, it is interesting to note that transformation could happen due to solely a hydrostatic stress state. Additionally, the lengths $\rho_{t_0}$ and $\rho_{c_0}$ -- representing the material's strength in tension and compression in a purely deviatoric stress state -- are shown in both \cref{fig:transformation-surface,fig:meridional-plane} and can be calculated as \citep{Chen1988}:
\begin{align}
  \rho_{t_0} & = \frac{2 \sqrt{6} c \cos \phi}{3 + \sin \phi} = \frac{2 \sqrt{6} \DGm}{3 \sqrt{\delta^2 + \xi^2} + \delta} = \SI{391.1}{\mega\pascal} \\
  \rho_{c_0} & = \frac{2 \sqrt{6} c \cos \phi}{3 - \sin \phi} = \frac{2 \sqrt{6} \DGm}{3 \sqrt{\delta^2 - \xi^2} + \delta} = \SI{414.5}{\mega\pascal},
\end{align}
which are once again easily verified to be in agreement with the results obtained. As the Mohr--Coulomb surface only needs two points to be fully characterised, these lengths provide a simple way of confirming the theoretical considerations that led to relating the martensitic transformation criterion to this particular yield criterion.

\FloatBarrier
\section{Conclusions} \label{sec:conclusions}

In this work, a recently proposed large-strain, multi-scale model for mechanically induced martensitic phase transformations is analysed. The model relies on the PTMC to obtain transformation properties such as shape change and habit plane orientations, also postulating a thermodynamically consistent generalisation of the stress-assisted transformation criterion of \citet{Patel1953}. It is shown that the model rigorously dissipates the mechanical energy barrier during the transformation course, in accord with experimental observations. The associated transformation functions also act as a variant selection mechanism: systems more favourably aligned with the local stress state are preferred for transformation. Particular attention is paid to the relation between this model's transformation criterion and the classical Mohr--Coulomb yield function. In fact, the transformation model is shown to be of a pressure-dependent nature, as a consequence of its non-isochoric character.

The constitutive model also includes the effects of austenite slip plasticity, incorporated using a finite-strain single crystal plasticity model. Even though the viscous properties are negligible for the alloys studied at the temperatures of interest, a visco-plastic formulation is used -- for both austenite plasticity and martensitic transformation -- to regularise the stress integration algorithm, preventing the difficulties associated to the selection of a set of active systems.

The mechanical behaviour of a meta-stable austenitic stainless steel is then analysed using this constitutive model. Given the extensive experimental data available in the literature, the chosen material provides a valuable test case for the constitutive model, as its meta-stable austenitic phase exhibits significant slip plasticity before the onset of the mechanically induced martensitic transformation. In a simple shear test, the overall effect of the martensitic transformation is captured by the model, but an overly stiff response is observed at higher strains. This effect is most likely due to the computational strategy adopted to handle the coupling between plasticity and transformation. A more robust formulation to address the simultaneous evolution of slip activity and transformation at intermediate martensite volume fractions is a topic of interest for further investigation. Nevertheless, the constitutive model shows particularly promising results in reproducing experimentally obtained transformation loci, providing further evidence that an energy-based transformation criterion can account for both the stress-assisted and the strain-induced transformation regimes.

\section*{Acknowledgements} The first and second authors gratefully acknowledge the Zienkiewicz Research Scholarships provided by Swansea University's College of Engineering during the course of much of this work.

\bibliography{\jobname}

\appendix
\section{Slip and transformation system vectors} \label{sec:appendix-systems}

\begin{table} [h]
  \caption{Octahedral slip systems in FCC single crystals.}
  \centering
  \begin{tabular}[c]{c c c | c c c}
    \rule{0pt}{0.8\normalbaselineskip}\hspace{-3pt}
    $\alpha$ & $\ma$                                    & $\sa$                                               & $\alpha$ & $\ma$                                    & $\sa$                        \\
    \hline\rule{0pt}{0.8\normalbaselineskip}\hspace{-3pt}
    1        & $\frac{1}{\sqrt{3}}[1, 1, 1]$            & $\frac{1}{\sqrt{2}}[\overline{1}, 0, 1]$            & 7        & $\frac{1}{\sqrt{3}}[\overline{1}, 1, 1]$ & $\frac{1}{\sqrt{2}}[0, \overline{1}, 1]$            \\[5pt]
    2        & $\frac{1}{\sqrt{3}}[1, 1, 1]$            & $\frac{1}{\sqrt{2}}[0, 1, \overline{1}]$            & 8        & $\frac{1}{\sqrt{3}}[\overline{1}, 1, 1]$ & $\frac{1}{\sqrt{2}}[\overline{1}, 0, \overline{1}]$            \\[5pt]
    3        & $\frac{1}{\sqrt{3}}[1, 1, 1]$            & $\frac{1}{\sqrt{2}}[1, \overline{1}, 0]$            & 9        & $\frac{1}{\sqrt{3}}[\overline{1}, 1, 1]$ & $\frac{1}{\sqrt{2}}[1, 1, 0]$                          \\[5pt]
    4        & $\frac{1}{\sqrt{3}}[1, \overline{1}, 1]$ & $\frac{1}{\sqrt{2}}[1, 0, \overline{1}]$            & 10       & $\frac{1}{\sqrt{3}}[1, 1, \overline{1}]$ & $\frac{1}{\sqrt{2}}[\overline{1}, 1, 0]$                       \\[5pt]
    5        & $\frac{1}{\sqrt{3}}[1, \overline{1}, 1]$ & $\frac{1}{\sqrt{2}}[\overline{1}, \overline{1}, 1]$ & 11       & $\frac{1}{\sqrt{3}}[1, 1, \overline{1}]$ & $\frac{1}{\sqrt{2}}[0, \overline{1}, \overline{1}]$ \\[5pt]
    6        & $\frac{1}{\sqrt{3}}[1, \overline{1}, 1]$ & $\frac{1}{\sqrt{2}}[0, 1, 1]$                       & 12       & $\frac{1}{\sqrt{3}}[1, 1, \overline{1}]$ & $\frac{1}{\sqrt{2}}[1, 0, 1]$                          \\[5pt]
  \end{tabular}
  \label{table:octahedral_systems}
\end{table}

\begin{table} [h]
  \caption{Transformation system vectors for the ASTM A-564 alloy. Numerical values:
    $m_1 \approx 0.608$, $m_2 \approx 0.178$, $m_3 \approx 0.774$, 
    $d_1 \approx 0.156$, $d_2 \approx 0.046$, $d_3 \approx 0.159$.}
  \centering
  \begin{tabular}[c]{c c c | c c c}
    \rule{0pt}{0.8\normalbaselineskip}\hspace{-3pt}
    $i$ & $\mi$                                   &  $\di$                                   & $i$ & $\mi$                                   &  $\di$                                  \\
    \hline\rule{0pt}{0.8\normalbaselineskip}\hspace{-3pt}
    1   & $[m_1, \overline{m_2}, m_3]$            &  $[\overline{d_1}, d_2, d_3]$            & 13  & $[\overline{m_2}, m_3, m_1]$            & $[d_2, d_3, \overline{d_1}]$            \\[2pt]
    2   & $[\overline{m_1}, m_2, m_3]$            &  $[d_1, \overline{d_2}, d_3]$            & 14  & $[m_2, m_3, \overline{m_1}]$            & $[\overline{d_2}, d_3, d_1]$            \\[2pt]
    3   & $[m_1, m_2, m_3]$                       &  $[\overline{d_1}, \overline{d_2}, d_3]$ & 15  & $[m_2, m_3, m_1]$                       & $[\overline{d_2}, d_3, \overline{d_1}]$ \\[2pt]
    4   & $[\overline{m_1}, \overline{m_2}, m_3]$ &  $[d_1, d_2, d_3]$                       & 16  & $[\overline{m_2}, m_3, \overline{m_1}]$ & $[d_2, d_3, d_1]$                       \\[2pt]
    5   & $[\overline{m_2}, m_1, m_3]$            &  $[d_2, \overline{d_1}, d_3]$            & 17  & $[m_3, m_1, \overline{m_2}]$            & $[d_3, \overline{d_1}, d_2]$            \\[2pt]
    6   & $[m_2, \overline{m_1}, m_3]$            &  $[\overline{d_2}, d_1, d_3]$            & 18  & $[m_3, \overline{m_1}, m_2]$            & $[d_3, d_1, \overline{d_2}]$            \\[2pt]
    7   & $[m_2, m_1, m_3]$                       &  $[\overline{d_2}, \overline{d_1}, d_3]$ & 19  & $[m_3, m_1, m_2]$                       & $[d_3, \overline{d_1}, \overline{d_2}]$ \\[2pt]
    8   & $[\overline{m_2}, \overline{m_1}, m_3]$ &  $[d_2, d_1, d_3]$                       & 20  & $[m_3, \overline{m_1}, \overline{m_2}]$ & $[d_3, d_1, d_2]$                       \\[2pt]
    9   & $[m_1, m_3, \overline{m_2}]$            &  $[\overline{d_1}, d_3, d_2]$            & 21  & $[m_3, \overline{m_2}, m_1]$            & $[d_3, d_2, \overline{d_1}]$            \\[2pt]
    10  & $[\overline{m_1}, m_3, m_2]$            &  $[d_1, d_3, \overline{d_2}]$            & 22  & $[m_3, m_2, \overline{m_1}]$            & $[d_3, \overline{d_2}, d_1]$            \\[2pt]
    11  & $[m_1, m_3, m_2]$                       &  $[\overline{d_1}, d_3, \overline{d_2}]$ & 23  & $[m_3, m_2, m_1]$                       & $[d_3, \overline{d_2}, \overline{d_1}]$ \\[2pt]
    12  & $[\overline{m_1}, m_3, \overline{m_2}]$ &  $[d_1, d_3, d_2]$                       & 24  & $[m_3, \overline{m_2}, \overline{m_1}]$ & $[d_3, d_2, d_1]$                       \\[2pt]
  \end{tabular}
  \label{table:transformation-systems}
\end{table}

\end{document}